\documentclass[11pt]{article}
\usepackage{amssymb,amsmath,amsfonts}
\usepackage{graphicx}
\usepackage{graphics}
\usepackage{eepic,epsfig}

\@addtoreset{equation}{section}

\makeatother

\begin{document}

\title{Casimir effect for scalar current densities \\
in topologically nontrivial spaces }
\author{ S. Bellucci$^{1}$\thanks{%
E-mail: bellucci@lnf.infn.it }, A. A. Saharian$^{2}$\thanks{%
E-mail: saharian@ysu.am }, N. A. Saharyan$^{2}$ \vspace{0.3cm} \\
\textit{$^1$ INFN, Laboratori Nazionali di Frascati,}\\
\textit{Via Enrico Fermi 40,00044 Frascati, Italy} \vspace{0.3cm}\\
\textit{$^2$ Department of Physics, Yerevan State University,}\\
\textit{1 Alex Manoogian Street, 0025 Yerevan, Armenia }}
\maketitle

\begin{abstract}
We evaluate the Hadamard function and the vacuum expectation value (VEV) of
the current density for a charged scalar field, induced by flat boundaries
in spacetimes with an arbitrary number of toroidally compactified spatial
dimensions. The field operator obeys the Robin conditions on the boundaries
and quasiperiodicity conditions with general phases along compact
dimensions. In addition, the presence of a constant gauge field is assumed.
The latter induces Aharonov-Bohm-type effect on the VEVs. There is a region
in the space of the parameters in Robin boundary conditions where the vacuum
state becomes unstable. The stability condition depends on the lengths of
compact dimensions and is less restrictive than that for background with
trivial topology. The vacuum current density is a periodic function of the
magnetic flux, enclosed by compact dimensions, with the period equal to the
flux quantum. It is explicitly decomposed into the boundary-free and
boundary-induced contributions. In sharp contrast to the VEVs of the field
squared and the energy-momentum tensor, the current density does not contain
surface divergences. Moreover, for Dirichlet condition it vanishes on the
boundaries. The normal derivative of the current density on the boundaries
vanish for both Dirichlet and Neumann conditions and is nonzero for general
Robin conditions. When the separation between the plates is smaller than
other length scales, the behavior of the current density is essentially
different for non-Neumann and Neumann boundary conditions. In the former
case, the total current density in the region between the plates tends to
zero. For Neumann boundary condition on both plates, the current density is
dominated by the interference part and is inversely proportional to the
separation.
\end{abstract}

\bigskip


\bigskip

\section{Introduction}

In a number of physical problems one needs to consider the model in the
background of manifolds with boundaries on which the dynamical variables
obey some prescribed boundary conditions. In quantum field theory, the
imposition of boundary conditions on the field operator gives rise to a
number of physical consequences. The Casimir effect is among the most
interesting phenomena of this kind (for reviews see \cite{Most97}). It
arises due to the modification of the quantum fluctuations of a field by
boundary conditions and plays an important role in different fields of
physics, from microworld to cosmology. The boundary conditions in the
Casimir effect may have different physical natures and can be divided into
two main classes. In the first one, the constraints are induced by the
presence of boundaries, like macroscopic bodies in QED, interfaces
separating different phases of a physical system, extended topological
defects, horizons in gravitational physics, branes in high-energy theories
with extra dimensions and in string theories. In the corresponding models
the field operator obeys the boundary condition on some spacelike surfaces
(static or dynamical). The original problem with two conducting plates,
discussed by Casimir in 1948 \cite{Casi48}, belongs to this class. Since the
original research by Casimir, many theoretical and experimental works have
been done on this problem for various types of bulk and boundary geometries.
Different methods have been developed including direct mode-summation and
the zeta function techniques, semiclassical methods, the optical approach,
worldline numerics, the path integral approach, methods based on scattering
theory, and numerical methods based on evaluation of the stress tensor via
the fluctuation-dissipation theorem. The recent high precision measurements
of the Casimir force allow for an accurate comparison between the
experimental results and theoretical predictions.

In the second class, the boundary conditions on the field operator are
induced by the nontrivial topology of the space. The changes in the
properties of the vacuum state generated by this type of conditions are
referred to as the topological Casimir effect. The importance of this effect
is motivated by that the presence of compact dimensions is an inherent
feature in many high-energy theories of fundamental physics, in cosmology
and in condensed matter physics. In particular, supergravity and superstring
theories are formulated in spacetimes having extra compact dimensions. The
compactified higher-dimensional models provide a possibility for the
unification of known interactions. Models of a compact universe with
nontrivial topology may also play an important role by providing proper
initial conditions for inflation in the early stages of the Universe
expansion \cite{Lind04}. In condensed matter physics, a number of planar
systems in the low-energy sector are described by an effective field theory.
The compactification of these systems leads to the change in the ground
state energy which is the analog of the topological Casimir effect. A
well-known example of this type of systems is a graphene sheet. In the long
wavelength limit, the dynamics of the quasiparticles for the electronic
subsystem is described in terms of the Dirac-like theory in two-dimensional
space (see Ref. \cite{Gusy07}). The corresponding effective 3-dimensional
relativistic field theory, in addition to Dirac fermions, involves scalar
and gauge fields (see \cite{Jack07} and references therein). The
single-walled carbon nanotubes are generated by rolling up a graphene sheet
to form a cylinder and for the corresponding Dirac model one has the spatial
topology $R^{1}\times S^{1}$. For another class of graphene-made structures,
called toroidal carbon nanotubes, the background topology is a 2-dimensional
torus, $T^{2}$.

Many authors have investigated the Casimir energies and stresses associated
with the presence of compact dimensions (for reviews see Refs. \cite%
{Most97,Duff86,Khan14}). In higher-dimensional models the Casimir energy of
bulk fields induces an effective potential for the compactification radius.
This has been used as a stabilization mechanism for the corresponding moduli
fields and as a source for dynamical compactification of the extra
dimensions during the cosmological evolution. The Casimir effect has also
been considered as a possible origin for the dark energy in both
Kaluza-Klein-type and braneworld models \cite{DarkEn}. Extra-dimensional
theories with low-energy compactification scale predict Yukawa-type
corrections to Newton's gravitational law and the measurements of the
Casimir forces between macroscopic bodies provide a sensitive test for
constraining the parameters of the corresponding long-range interactions
\cite{Most87}. The influence of extra compactified dimensions on the Casimir
effect in the classical configuration of two parallel plates has been
recently discussed for scalar \cite{Chen06}, electromagnetic \cite{Popp04}
and fermionic \cite{Bell09} fields.

The vast majority of the works on the influence of the copmactification on
the properties of the quantum vacuum in the Casimir effect has been
concerned with global quantities such as the force or the total energy. More
detailed information on the vacuum fluctuations is contained in the local
characteristics. Among the most important local quantities, because of their
close connection with the structure of spacetime, are the vacuum expectation
values (VEVs) of the vacuum energy density and stresses. For charged fields,
another important characteristic is the VEV of the current density. Due to
the global nature of the vacuum, this VEV carries information on both global
and local properties of the vacuum state. Besides, the VEV of the current
density appears as a source of the electromagnetic field in semiclassical
Maxwell equations, and, hence, it is needed in modeling a self-consistent
dynamics involving the electromagnetic field.

In models with nontrivial topology, the nonzero current densities in the
vacuum state may appear as a consequence of quasiperiodicity conditions
along compact dimensions or by the presence of gauge field fluxes enclosed
by these dimensions. Note that the gauge field fluxes in higher-dimensional
models will also generate a potential for moduli fields and this provides
another mechanism for moduli stabilization (for a review see \cite{Doug07}).
The VEV of the fermionic current density in spaces with toroidally
compactified dimensions has been considered in \cite{Bell10}. In the special
case of a 2-dimensional space, application are given to the electrons in
cylindrical and toroidal carbon nanotubes, described within the framework of
the effective field theory in terms of Dirac fermions. The vacuum currents
for charged fields in de Sitter and anti-de Sitter spacetimes with
toroidally compact spatial dimensions are investigated in \cite%
{Bell13,Beze15}. Finite temperature effects on the charge density and on the
current densities along compact dimensions have been discussed in \cite%
{Beze13} and \cite{Bell14} for scalar and fermionic fields, respectively.
The changes in the fermionic vacuum currents induced by the presence of
parallel plane boundaries, with the bag boundary conditions on them, are
investigated in \cite{Bell13b}.

In the present paper we consider the effect of two parallel plane boundaries
on the vacuum expectation value of the current density for a charged scalar
field in background spacetime with spatial topology $R^{p+1}\times T^{q}$,
where $T^{q}$ stands for a $q$-dimensional torus. The organization of the
paper is as follows. In the next section the geometry of the problem is
described and the Hadamard function is evaluated in the region between the
plates for general Robin boundary conditions. By using the expression for
the Hadamard function, in Section \ref{sec:1Plate}, we evaluate the current
density in the geometry of a single plate. The corresponding asymptotics are
discussed in various limiting cases and numerical results are presented. In
Section \ref{sec:2Plates} the current density is investigated in the region
between two plates. The main results of the paper are summarized in Section %
\ref{Sec:Conc}. An alternative representation of the Hadamard function is
given in Appendix.

\section{Formulation of the problem and the Hadamard function}

\label{Sec:Hadam}

We consider $(D+1)$-dimensional flat spacetime with spatial topology $%
R^{p+1}\times T^{q}$, $p+q+1=D$ (for a review of quantum field-theoretical
effects in toroidal topology see Ref. \cite{Khan14}). The set of Cartesian
coordinates in the subspace $R^{p+1}$ will be denoted by $\mathbf{x}%
_{p+1}=(x^{1},...,x^{p+1})$ and the corresponding coordinates on the torus
by $\mathbf{x}_{q}=(x^{p+2},...,x^{D})$. If $L_{l}$ is the length of the $l$%
th compact dimension then one has $-\infty <x^{l}<\infty $ for $l=1,..,p$,
and $0\leqslant x^{l}\leqslant L_{l}$ for $l=p+2,...,D$. Our main interest
in this paper is the VEV of the current density for a quantum scalar field $%
\varphi (x)$ with the mass $m$ and charge $e$. The equation for the field
operator reads%
\begin{equation}
\left( g^{\mu \nu }D_{\mu }D_{\nu }+m^{2}\right) \varphi =0,  \label{Feq}
\end{equation}%
where $g^{\mu \nu }=\mathrm{diag}(1,-1,\ldots ,-1)$, $D_{\mu }=\partial
_{\mu }+ieA_{\mu }$ and $A_{\mu }$ is the vector potential for a classical
gauge field. We assume the presence of two parallel flat boundaries\footnote{%
In analogy with the standard Casimir effect, in the discussion below we will
refer the boundaries as plates.} placed at $x^{p+1}=a_{1}$ and $x^{p+1}=a_{2}
$, on which the field obeys Robin boundary conditions
\begin{equation}
(1+\beta _{j}n_{j}^{\mu }D_{\mu })\varphi (x)=0,\quad x^{p+1}\equiv z=a_{j},
\label{Rob}
\end{equation}%
with constant coefficients $\beta _{j}$, $j=1,2$, and with $n_{j}^{\mu }$
being the inward pointing normal to the boundary at $x^{p+1}=a_{j}$. Here,
for the further convenience we have introduced a special notation $z=x^{p+1}$
for the $(p+1)$th spatial dimension. Note that Robin boundary conditions in
the form (\ref{Rob}) are gauge invariant (for the discussion of various
types of gauge invariant boundary conditions see \cite{Espo97}). In what
follows we will consider the region between the plates, $a_{1}\leqslant
z\leqslant a_{2}$. For this region one has $n_{j}^{\mu }=(-1)^{j-1}\delta
_{p+1}^{\mu }$. The expressions for the VEVs in the regions $z\leqslant a_{1}
$ and $z\geqslant a_{2}$ are obtained by the limiting transitions. The
results for Dirichlet and Neumann boundary conditions are obtained from
those for the condition (\ref{Rob}) in the limits $\beta _{j}\rightarrow 0$
and $\beta _{j}\rightarrow \infty $, $A_{\mu }=0$, respectively. Robin type
conditions appear in a variety of situations, including the considerations
of vacuum effects for a confined charged scalar field in external fields
\cite{Ambj83}, gauge field theories, quantum gravity and supergravity \cite%
{Espo97,Luck91}, braneworld models \cite{Gher00} and in a class of models
with boundaries separating the spatial regions with different gravitational
backgrounds \cite{Bell14b}. In some geometries, these conditions may be
useful for depicting the finite penetration of the field into the boundary
with the "skin-depth" parameter related to the coefficient $\beta _{j}$. It
is interesting to note that the quantum scalar field constrained by Robin
condition on the boundary of cavity violates the Bekenstein's
entropy-to-energy bound near certain points in the space of the parameter $%
\beta _{j}$ \cite{Solo01}.

In addition to the boundary conditions on the plates, for the theory to be
completely defined, we should also specify the periodicity conditions along
the compact dimensions. Different conditions correspond to topologically
inequivalent field configurations \cite{Isha78}. Here, we consider generic
quasiperiodicity conditions,%
\begin{equation}
\varphi (t,x^{1},\ldots ,x^{l}+L_{l},\ldots ,x^{D})=e^{i\alpha _{l}}\varphi
(t,x^{1},\ldots ,x^{l},\ldots ,x^{D}),  \label{PerCond}
\end{equation}%
with constant phases $\alpha _{l}$, $l=p+2,\ldots ,D$. The special cases of
the condition (\ref{PerCond}) with $\alpha _{l}=0$ and $\alpha _{l}=\pi $
correspond to the most frequently discussed cases of untwisted and twisted
scalar fields, respectively. As it will be seen below, one of the effects of
nontrivial phases in (\ref{PerCond}) is the appearance of nonzero vacuum
currents along compact dimensions (for a discussion of physical effects of
phases in periodicity conditions along compact dimensions see \cite{Sche79}
and references therein).

For a scalar field, the operator of the current density is given by the
expression
\begin{equation}
j_{\mu }(x)=ie[\varphi ^{+}(x)D_{\mu }\varphi (x)-(D_{\mu }\varphi
(x))^{+}\varphi (x)],  \label{jl}
\end{equation}%
$l=0,1,\ldots ,D$. Its VEV\ is obtained from the Hadamard function
\begin{equation}
G(x,x^{\prime })=\langle 0|\varphi (x)\varphi ^{+}(x^{\prime })+\varphi
^{+}(x^{\prime })\varphi (x)|0\rangle ,  \label{G1}
\end{equation}%
with $|0\rangle $ being the vacuum state, by using the formula
\begin{equation}
\langle 0|j_{\mu }(x)|0\rangle \equiv \langle j_{\mu }(x)\rangle =\frac{i}{2}%
e\lim_{x^{\prime }\rightarrow x}(\partial _{\mu }-\partial _{\mu }^{\prime
}+2ieA_{\mu })G(x,x^{\prime }).  \label{jl1}
\end{equation}

In the discussion below we will assume a constant gauge field $A_{\mu }$.
Though the corresponding field strength vanishes, the nontrivial topology of
the background spacetime leads to the Aharonov-Bohm-like effects on physical
observables. In the case of a constant gauge field $A_{\mu }$, the latter
can be excluded from the field equation and from the expression for the VEV
of the current density by the gauge transformation $A_{\mu }=A_{\mu
}^{\prime }+\partial _{\mu }\chi $, $\varphi (x)=e^{-ie\chi }\varphi
^{\prime }(x)$, with the function $\chi =A_{\mu }x^{\mu }$. In the new gauge
one has $A_{\mu }^{\prime }=0$. However, unlike to the case of trivial
topology, here the constant vector potential does not completely disappear
from the problem. It appears in the periodicity conditions for the new field
operator:%
\begin{equation}
\varphi ^{\prime }(t,x^{1},\ldots ,x^{l}+L_{l},\ldots ,x^{D})=e^{i\tilde{%
\alpha}_{l}}\varphi ^{\prime }(t,x^{1},\ldots ,x^{l},\ldots ,x^{D}),
\label{PerCond2}
\end{equation}%
where now the phases are given by the expression%
\begin{equation}
\tilde{\alpha}_{l}=\alpha _{l}+eA_{l}L_{l}.  \label{alftilde}
\end{equation}%
In the discussion below we shall consider the problem in the gauge $(\varphi
^{\prime }(x),A_{\mu }^{\prime }=0)$ omitting the prime. For this gauge, in (%
\ref{Feq}), (\ref{Rob}), (\ref{jl}) one has $D_{\mu }=\partial _{\mu }$ and
in the expressions (\ref{jl1}) the term with the vector potential is absent.

From the discussion above it follows that in the problem at hand the
presence of a constant gauge field is equivalent to the shift in the phases
of the periodicity conditions along compact dimensions. The shift in the
phase is expressed in terms of the magnetic flux $\Phi _{l}$ enclosed by the
$l$th compact dimension as%
\begin{equation}
eA_{l}L_{l}=-e\mathbf{A}_{l}L_{l}=-2\pi \Phi _{l}/\Phi _{0},  \label{Flux}
\end{equation}%
where $\Phi _{0}=2\pi /e$ is the flux quantum and $\mathbf{A}_{l}$ is the $l$%
th component of the spatial vector $\mathbf{A}=(-A_{1},\ldots ,-A_{D})$. In
the discussion below the physical effects of a constant gauge field will
appear through the phases $\tilde{\alpha}_{l}$. In particular, the VEVs of
physical observables are periodic functions of these phases with the period $%
2\pi $. In terms of the magnetic flux, this corresponds to the periodicity
of the VEVs, as functions of the magnetic flux, with the period equal to the
flux quantum.

For the evaluation of the Hadamard function in (\ref{jl1}) we shall use the
mode-sum formula%
\begin{equation}
G(x,x^{\prime })=\sum_{\mathbf{k}}\sum_{s=\pm }\varphi _{\mathbf{k}%
}^{(s)}(x)\varphi _{\mathbf{k}}^{(s)\ast }(x^{\prime }),  \label{G1n}
\end{equation}%
where $\varphi _{\mathbf{k}}^{(\pm )}(x)$ form a complete set of normalised
positive- and negative-energy solutions to the classical field equation
obeying the boundary conditions of the model. In the region between the
plates, introducing the wave vectors $\mathbf{k}_{p}=(k_{1},\ldots ,k_{p})$
and $\mathbf{k}_{q}=(k_{p+2},\ldots ,k_{D})$, these mode functions can be
written in the form%
\begin{equation}
\varphi _{\mathbf{k}}^{(\pm )}(x)=C_{\mathbf{k}}\cos \left[ k_{p+1}\left(
z-a_{j}\right) +\gamma _{j}(k_{p+1})\right] e^{i\mathbf{k}_{\parallel }\cdot
\mathbf{x}_{\parallel }\mp i\omega _{\mathbf{k}}t},  \label{phik}
\end{equation}%
where $\mathbf{k}_{\parallel }=(\mathbf{k}_{p},\mathbf{k}_{q})$, $\mathbf{k}%
=(\mathbf{k}_{p},k_{p+1},\mathbf{k}_{q})$, $\omega _{\mathbf{k}}=\sqrt{%
\mathbf{k}^{2}+m^{2}}$, and $\mathbf{x}_{\parallel }$ stands for the
coordinates parallel to the plates. For the momentum components along the
dimensions $x^{i}$, $i=1,\ldots ,p$, one has $-\infty <k_{i}<+\infty $,
whereas the components along the compact dimensions are quantized by the
periodicity conditions (\ref{PerCond2}):%
\begin{equation}
k_{l}=\left( 2\pi n_{l}+\tilde{\alpha}_{l}\right) /L_{l},\quad n_{l}=0,\pm
1,\pm 2,\ldots .,  \label{kltild}
\end{equation}%
with $l=p+2,...,D$. We will denote by $\omega _{0}$ the smallest value for
the energy in the compact subspace, $\sqrt{\mathbf{k}_{q}^{2}+m^{2}}%
\geqslant \omega _{0}$. Assuming that $|\tilde{\alpha}_{l}|\leqslant \pi $,
we have%
\begin{equation}
\omega _{0}=\sqrt{\sum\nolimits_{l=p+2}^{D}\tilde{\alpha}%
_{l}^{2}/L_{l}^{2}+m^{2}}.  \label{om0}
\end{equation}%
This quantity can be considered as the effective mass for the field quanta.

Now we should impose on the modes (\ref{phik}) the boundary conditions (\ref%
{Rob}) with $D_{\mu }=\partial _{\mu }$. From the boundary condition on the
plate at $z=a_{j}$, for the function $\gamma _{j}(k_{p+1})$ in (\ref{phik})
one gets
\begin{equation}
e^{2i\gamma _{j}(k_{p+1})}=\frac{ik_{p+1}\beta _{j}(-1)^{j}+1}{ik_{p+1}\beta
_{j}(-1)^{j}-1}.  \label{alfj}
\end{equation}%
From the boundary condition on the second plate it follows that the
eigenvalues for $k_{p+1}$ are solutions of the equation
\begin{equation}
e^{2iy}=\frac{1+ib_{2}y}{1-ib_{2}y}\frac{1+ib_{1}y}{1-ib_{1}y},
\label{EigEq}
\end{equation}%
where
\begin{equation}
y=k_{p+1}a,\;b_{j}=\beta _{j}/a,  \label{bj}
\end{equation}%
and $a=a_{2}-a_{1}$ is the separation between the plates. Formula (\ref%
{EigEq}) can also be written in the form%
\begin{equation}
\left( 1-b_{1}b_{2}y^{2}\right) \sin y-(b_{2}+b_{1})y\cos y=0.
\label{EigEq2}
\end{equation}%
Unlike to the cases of Dirichlet and Neumann conditions, for Robin boundary
condition the eigenvalues of $k_{p+1}$ are given implicitly, as solutions of
the transcendental equation (\ref{EigEq2}). This equation has an infinite
number of positive roots which will be denoted by $y=\lambda _{n}$, $%
n=1,2,\ldots $, and for the corresponding eigenvalues of $k_{p+1}$ one has $%
k_{p+1}=\lambda _{n}/a$. For $b_{j}\leqslant 0$ or $\{b_{1}+b_{2}\geqslant
1,b_{1}b_{2}\leqslant 0\}$ there are no other roots in the right-half plane
of a complex variable $y$, $\mathrm{Re}\,y\geqslant 0$ (see \cite{Rome02}).
In the remaining region of the plane $(b_{1},b_{2})$, the equation (\ref%
{EigEq2}) has purely imaginary roots $\pm iy_{l}$, $y_{l}>0$. Depending on
the values of $b_{j}$, the number of $y_{l}$ can be one or two. In the
presence of purely imaginary roots, under the condition $\omega _{0}<y_{l}$,
there are modes of the field for which the energy $\omega _{\mathbf{k}}$
becomes imaginary. This would lead to the instability of the vacuum state.
In the discussion below we will assume that $\omega _{0}>y_{l}$. Note that
in the corresponding problem on background of spacetime with trivial
topology the stability condition is written as $m>y_{l}$. Now, by taking
into account that $\omega _{0}>m$, we conclude that the compactification, in
general, enlarges the stability range in the space of parameters of Robin
boundary conditions.

The coefficient $C_{\mathbf{k}}$ in (\ref{phik}) is found from the
orthonormalization condition
\begin{equation}
\int d^{D}x\varphi _{\mathbf{k}}^{(\lambda )}(x)\varphi _{\mathbf{k}^{\prime
}}^{(\lambda ^{\prime })\ast }(x)=\frac{\delta _{\lambda \lambda ^{\prime }}%
}{2\omega _{\mathbf{k}}}\delta (\mathbf{k}_{p}-\mathbf{k}_{p}^{\prime
})\delta _{nn^{\prime }}\delta _{n_{p+2},n_{p+2}^{\prime }}....\delta
_{n_{D},n_{D}^{\prime }},  \label{Norm}
\end{equation}%
where the integration over $x^{p+1}$ goes in the region between the plates.
Substituting the functions (\ref{phik}), one gets%
\begin{equation}
|C_{\mathbf{k}}|^{2}=\frac{\left\{ 1+\cos [y+2\tilde{\gamma}_{j}(y)]\sin
(y)/y\right\} ^{-1}}{(2\pi )^{p}aV_{q}\omega _{\mathbf{k}}},  \label{Ck}
\end{equation}%
where $y$ is a root of the equation (\ref{EigEq2}) and $%
V_{q}=L_{p+1}....L_{D}$ is the volume of the compact subspace. The function $%
\tilde{\gamma}_{j}(y)$ is defined by the relation%
\begin{equation}
e^{2i\tilde{\gamma}_{j}(y)}=\frac{iyb_{j}-1}{iyb_{j}+1}.  \label{alfjtild}
\end{equation}%
First we shall consider the case when all the roots of (\ref{EigEq2}) are
real and $y=\lambda _{n}$.

Having the complete set of normalized mode functions, the mode-sum (\ref{G1n}%
) for the Hadamard function is written in the form%
\begin{eqnarray}
G(x,x^{\prime }) &=&\frac{1}{aV_{q}}\int \frac{d\mathbf{k}_{p}}{(2\pi )^{p}}%
\sum_{\mathbf{n}_{q}}\sum_{n=1}^{\infty }\frac{1}{\omega _{\mathbf{k}}}%
g_{j}(z,z^{\prime },\lambda _{n}/a)  \notag \\
&&\times \frac{\lambda _{n}\cos (\omega _{\mathbf{k}}\Delta t)e^{i\mathbf{k}%
_{p}\cdot \Delta \mathbf{x}_{p}+i\mathbf{k}_{q}\cdot \Delta \mathbf{x}_{q}}}{%
\lambda _{n}+\cos \left[ \lambda _{n}+2\tilde{\gamma}_{j}(\lambda _{n})%
\right] \sin \lambda _{n}},  \label{G11}
\end{eqnarray}%
where $\Delta \mathbf{x}_{p}\mathbf{=x}_{p}-\mathbf{x}_{p}^{\prime }$, $%
\Delta \mathbf{x}_{q}\mathbf{=x}_{q}-\mathbf{x}_{q}^{\prime }$, $\Delta
t=t-t^{\prime }$, and $\mathbf{n}_{q}=(n_{p+2},\ldots ,n_{D})$, $-\infty
<n_{l}<+\infty $. In (\ref{G11}), the energy for the mode with a given $%
\mathbf{k}$ is written as%
\begin{equation}
\omega _{\mathbf{k}}=\sqrt{\mathbf{k}_{p}^{2}+\lambda _{n}^{2}/a^{2}+\omega
_{\mathbf{n}_{q}}^{2}},  \label{omk}
\end{equation}%
and%
\begin{equation}
\omega _{\mathbf{n}_{q}}=\sqrt{\mathbf{k}_{q}^{2}+m^{2}},\;\mathbf{k}%
_{q}^{2}=\sum_{l=p+2}^{D}\left( \frac{2\pi n_{l}+\tilde{\alpha}_{l}}{L_{l}}%
\right) ^{2}.  \label{omnq}
\end{equation}%
Here and in what follows we use the notation%
\begin{equation}
g_{j}(z,z^{\prime },u)=\cos \left( u\Delta z\right) +\frac{1}{2}\sum_{s=\pm
1}e^{siy|z+z^{\prime }-2a_{j}|}\frac{iu\beta _{j}-s}{iu\beta _{j}+s}.
\label{gj}
\end{equation}%
Note that $g_{j}(z,z^{\prime },-y)=g_{j}(z,z^{\prime },y)$ and $%
g_{j}(z,z^{\prime },0)=0$.

In (\ref{G11}), the eigenvalues $\lambda _{n}$ are given implicitly and this
expression is not convenient for the evaluation of the VEVs. In order to
obtain an expression in which the explicit knowledge of $\lambda _{n}$ is
not required, we apply to the series over $n$ the Abel-Plana-type summation
formula \cite{Rome02,Saha08Rev}%
\begin{eqnarray}
\sum_{n=1}^{\infty }\frac{\pi \lambda _{n}f(\lambda _{n})}{\lambda _{n}+\cos
[\lambda _{n}+2\tilde{\gamma}_{j}(\lambda _{n})]\sin \lambda _{n}} &=&-\frac{%
\pi f(0)/2}{1-b_{2}-b_{1}}+\int_{0}^{\infty }duf(u)  \notag \\
&&+i\int_{0}^{\infty }du\frac{f(iu)-f(-iu)}{c_{1}(u)c_{2}(u)e^{2u}-1},
\label{sumfor}
\end{eqnarray}%
where, for the further convenience, the notation%
\begin{equation}
c_{j}(u)=\frac{b_{j}u-1}{b_{j}u+1}  \label{cj}
\end{equation}%
is introduced. In (\ref{sumfor}) we have assumed that $b_{j}\leqslant 0$.
The changes in the evaluation procedure in the case $b_{j}>0$ will be
discussed below. For the series in (\ref{G11}), we take in the summation
formula%
\begin{equation}
f(\lambda _{n})=\frac{\cos (\omega _{\mathbf{k}}\Delta t)}{\omega _{\mathbf{k%
}}}g_{j}(z,z^{\prime },\lambda _{n}/a).  \label{flamb}
\end{equation}%
Note that $f(0)=0$ and the first term in the right-hand side of (\ref{sumfor}%
) is absent.

The use of the summation formula (\ref{sumfor}) with (\ref{flamb}) allows us
to write the Hadamard function in the decomposed form%
\begin{eqnarray}
G(x,x^{\prime }) &=&G_{j}(x,x^{\prime })+\frac{2}{\pi V_{q}}\int \frac{d%
\mathbf{k}_{p}}{(2\pi )^{p}}\sum_{\mathbf{n}_{q}}\int_{a\omega _{\mathbf{k}%
_{\parallel }}}^{\infty }du\,g_{j}(z,z^{\prime },iu/a)  \notag \\
&&\times \frac{e^{i\mathbf{k}_{p}\cdot \Delta \mathbf{x}_{p}+i\mathbf{k}%
_{q}\cdot \Delta \mathbf{x}_{q}}}{c_{1}(u)c_{2}(u)e^{2u}-1}\frac{\cosh
(\Delta t\sqrt{u^{2}/a^{2}-\omega _{\mathbf{k}_{\parallel }}})}{\sqrt{%
u^{2}-a^{2}\omega _{\mathbf{k}_{\parallel }}}},  \label{G(1)}
\end{eqnarray}%
where $\omega _{\mathbf{k}_{\parallel }}=\sqrt{\mathbf{k}_{p}^{2}+\omega _{%
\mathbf{n}_{q}}^{2}}$. Here, the part%
\begin{eqnarray}
G_{j}(x,x^{\prime }) &=&\frac{1}{\pi V_{q}}\int \frac{d\mathbf{k}_{p}}{(2\pi
)^{p}}\sum_{\mathbf{n}_{q}}e^{i\mathbf{k}_{p}\cdot \Delta \mathbf{x}_{p}+i%
\mathbf{k}_{q}\cdot \Delta \mathbf{x}_{q}}  \notag \\
&&\times \int_{0}^{\infty }dk_{p+1}\frac{\cos (\omega _{\mathbf{k}}\Delta t)%
}{\omega _{\mathbf{k}}}g_{j}(z,z^{\prime },k_{p+1}),  \label{Gj}
\end{eqnarray}%
comes from the first integral in the right-hand side of (\ref{sumfor}) and
corresponds to the Hadamard function in the geometry of a single plate at $%
x^{p+1}=a_{j}$ when the second plate is absent. This function is further
decomposed by taking into account that the part in (\ref{gj}) coming from
the first term in the right-hand side of (\ref{gj}),
\begin{equation}
G_{0}(x,x^{\prime })=\frac{1}{V_{q}}\int \frac{d\mathbf{k}_{p+1}}{(2\pi
)^{p+1}}\sum_{\mathbf{n}_{q}}e^{i\mathbf{k}_{p+1}\cdot \Delta \mathbf{x}%
_{p+1}+i\mathbf{k}_{q}\cdot \Delta \mathbf{x}_{q}}\frac{\cos (\omega _{%
\mathbf{k}}\Delta t)}{\omega _{\mathbf{k}}},  \label{G0}
\end{equation}%
is the Hadamard function for the boundary-free geometry. After the
integration over the components of the momentum along uncompactified
dimensions, this function can be presented in the form%
\begin{equation}
G_{0}(x,x^{\prime })=\frac{2V_{q}^{-1}}{(2\pi )^{p/2+1}}\sum_{\mathbf{n}%
_{q}}e^{i\mathbf{k}_{q}\cdot \Delta \mathbf{x}_{q}}\omega _{\mathbf{n}%
_{q}}^{p}f_{p/2}(\omega _{\mathbf{n}_{q}}\sqrt{|\Delta \mathbf{x}%
_{p+1}|^{2}-(\Delta t)^{2}}),  \label{G0b}
\end{equation}%
with the notations
\begin{equation}
f_{\nu }(x)=K_{\nu }(x)/x^{\nu },  \label{fnu}
\end{equation}%
where $K_{\nu }(x)$ is the Macdonald function.

Consequently, the Hadamard function in the geometry of a single plate is
written as%
\begin{eqnarray}
G_{j}(x,x^{\prime }) &=&G_{0}(x,x^{\prime })+\frac{1}{2\pi V_{q}}\int \frac{d%
\mathbf{k}_{p}}{(2\pi )^{p}}\sum_{\mathbf{n}_{q}}e^{i\mathbf{k}_{p}\cdot
\Delta \mathbf{x}_{p}+i\mathbf{k}_{q}\cdot \Delta \mathbf{x}_{q}}  \notag \\
&&\times \sum_{s=\pm 1}\int_{0}^{\infty }dk_{p+1}\frac{\cos (\omega _{%
\mathbf{k}}\Delta t)}{\omega _{\mathbf{k}}}e^{sik_{p+1}|z+z^{\prime
}-2a_{j}|}\frac{ik_{p+1}\beta _{j}-s}{ik_{p+1}\beta _{j}+s},  \label{G1j}
\end{eqnarray}%
where the second term in the right-hand side is induced by the presence of
the plate at $x^{p+1}=a_{j}$. For the further transformation of the
boundary-induced part in (\ref{G1j}) we rotate the integration contour over $%
k_{p+1}$ by the angle $s\pi /2$. In the summation over $s$ the integrals
over the intervals $(0,\pm i\omega _{\mathbf{k}_{\parallel }})$ cancel each
other and we get%
\begin{eqnarray}
G_{j}(x,x^{\prime }) &=&G_{0}(x,x^{\prime })+\frac{1}{\pi V_{q}}\int \frac{d%
\mathbf{k}_{p}}{(2\pi )^{p}}\sum_{\mathbf{n}_{q}}e^{i\mathbf{k}_{p}\cdot
\Delta \mathbf{x}_{p}+i\mathbf{k}_{q}\cdot \Delta \mathbf{x}_{q}}  \notag \\
&&\times \int_{\omega _{\mathbf{k}_{\parallel }}}^{\infty }du\frac{\cosh
(\Delta t\sqrt{u^{2}-\omega _{\mathbf{k}_{\parallel }}^{2}})}{\sqrt{%
u^{2}-\omega _{\mathbf{k}_{\parallel }}^{2}}}\frac{u\beta _{j}+1}{u\beta
_{j}-1}e^{-u|z+z^{\prime }-2a_{j}|}.  \label{G1jb}
\end{eqnarray}%
This expression is well suited for the investigation of the current density.
With the representation (\ref{G1jb}), the Hadamard function in the region
between the plates, given by (\ref{G(1)}), is decomposed into the
boundary-free, single plate-induced and second plate-induced contributions.
An alternative expression for the Hadamard function is obtained in Appendix.

In deriving (\ref{G(1)}) and (\ref{G1jb}) we have assumed that $\beta
_{j}\leqslant 0$. In the case $\beta _{j}>0$, the quantum scalar field in
the geometry of a single plate at $z=a_{j}$ has modes with $k_{p+1}=i/\beta
_{j}$ for which the dependence on the coordinate $x^{p+1}$ has the form $%
e^{-z_{j}/\beta _{j}}$. In the case $1/\beta _{j}>\omega _{0}$, for a part
of these modes the energy is imaginary and the vacuum is unstable. In order
to have a stable vacuum, in what follows, for non-Dirichlet boundary
conditions, we shall assume that $1/\beta _{j}<\omega _{0}$ and the mode
with $k_{p+1}=i/\beta _{j}$ corresponds to a bound state. For $\beta _{j}>0$
and in the absence of purely imaginary roots of (\ref{EigEq2}), in the
right-hand side of the summation formula (\ref{sumfor}) the residue terms at
$u=\pm i/b_{j}$ should be added (see \cite{Rome02}). Now the integrand in (%
\ref{G1j}) has a simple pole at $k_{p+1}=is/\beta _{j}$ and after the
rotation the contribution of the residue at that pole should be added. This
contribution cancels the additional residue term in the right-hand side of (%
\ref{sumfor}). In the case when the equation (\ref{EigEq2}) has purely
imaginary roots the corresponding contributions have to be added to the
mode-sum (\ref{G11}) for the Hadamard function. But the corresponding
contributions should also be added in the left-hand side of (\ref{sumfor})
and the further evaluation procedure remains the same. Hence, the
expressions (\ref{G(1)}) and (\ref{G1jb}) are valid for all values of the
coefficients in the Robin boundary conditions. The only restrictions come
from the stability of the vacuum state: $1/\beta _{j}<\omega _{0}$ and $%
y_{l}<\omega _{0}$. In the presence of compact dimensions with $\tilde{\alpha%
}_{l}\neq 0$ one has $\omega _{0}>m$ and these conditions are less
restrictive than those in the case of trivial topology.

The current density in the boundary-free geometry is obtained by using the
Hadamard function (\ref{G0b}) and has been investigated in \cite{Beze13}.
The corresponding charge density and the current densities along uncompact
dimensions vanish. As it can be seen from (\ref{G(1)}) and (\ref{G1jb}), the
same holds in the case of the boundary-induced contributions in the VEVs.
Hence, the only nonzero components correspond to the current density along
compact dimensions.

\section{Vacuum currents in the geometry of a single plate}

\label{sec:1Plate}

In this section we investigate the VEV of the vacuum current density in the
geometry of a single plate at $x^{p+1}=a_{j}$. This VEV is obtained with the
help of the formula (\ref{jl1}) by using the Hadamard function from (\ref%
{G1jb}). The component of the VEV of the current density along the $l$th
compact dimension is presented in the decomposed form%
\begin{equation}
\langle j^{l}\rangle _{j}=\langle j^{l}\rangle _{0}+\langle j^{l}\rangle
_{j}^{(1)},  \label{jjdec}
\end{equation}%
where $\langle j^{l}\rangle _{0}$ is the current density in the
boundary-free geometry and $\langle j^{l}\rangle _{j}^{(1)}$ is the
contribution induced by the presence of the plate.

The current density in the boundary-free geometry has been investigated in
\cite{Beze13} and for the completeness we will recall the main results. The
current density is given by the formula%
\begin{eqnarray}
\langle j^{l}\rangle _{0} &=&\frac{4eL_{l}m^{D+1}}{(2\pi )^{(D+1)/2}}%
\sum_{n_{l}=1}^{\infty }n_{l}\sin (n_{l}\tilde{\alpha}_{l})  \notag \\
&&\times \sum_{\mathbf{n}_{q-1}}\cos (\mathbf{n}_{q-1}\cdot \tilde{%
\boldsymbol{\alpha }}_{q-1})f_{\frac{D+1}{2}}(mg_{\mathbf{n}_{q}}(\mathbf{L}%
_{q})),  \label{jl0a}
\end{eqnarray}%
where $\tilde{\boldsymbol{\alpha }}_{q-1}=(\tilde{\alpha}_{p+2},\ldots ,%
\tilde{\alpha}_{l-1},\tilde{\alpha}_{l+1},\ldots ,\tilde{\alpha}_{D})$, $%
\mathbf{n}_{q-1}=(n_{p+2},\ldots ,n_{l-1},n_{l+1},\ldots ,n_{D})$, and $g_{%
\mathbf{n}_{q}}(\mathbf{L}_{q})=(\sum_{i=p+2}^{D}n_{i}^{2}L_{i}^{2})^{1/2}$.
The current density $\langle j^{l}\rangle _{0}$ is an odd periodic function
of $\tilde{\alpha}_{l}$ with the period $2\pi $ and an even periodic
function of $\tilde{\alpha}_{r}$, $r\neq l$, with the same period. This
corresponds to the periodicity in the magnetic flux with the period of flux
quantum. An alternative expression for the current density in the
boundary-free geometry is given by the formula%
\begin{equation}
\langle j^{l}\rangle _{0}=\frac{4eL_{l}/V_{q}}{(2\pi )^{(p+3)/2}}%
\sum_{n=1}^{\infty }\frac{\sin \left( n\tilde{\alpha}_{l}\right) }{%
(nL_{l})^{p+2}}\sum_{\mathbf{n}_{q-1}}g_{\frac{p+3}{2}}(nL_{l}\omega _{%
\mathbf{n}_{q-1}}),  \label{jl0b}
\end{equation}%
where we have defined the function
\begin{equation}
g_{\nu }(x)=x^{\nu }K_{\nu }(x),  \label{gnu}
\end{equation}%
and%
\begin{equation}
\omega _{\mathbf{n}_{q-1}}^{2}=\omega _{\mathbf{n}_{q}}^{2}-k_{l}^{2}.
\label{omnq-1}
\end{equation}%
In the model with a single compact dimension ($q=1$) the representations (%
\ref{jl0a}) and (\ref{jl0b}) are identical.

When the length of the $l$th compact dimension, $L_{l}$, is much larger than
the other length scales, the behavior of the current density crucially
depends whether the parameter%
\begin{equation}
\omega _{0l}=\left( \sum\nolimits_{i=p+2,\neq l}^{D}\tilde{\alpha}%
_{i}^{2}/L_{i}^{2}+m^{2}\right) ^{1/2},  \label{om0l}
\end{equation}%
is zero or not. For $\omega _{0l}=0$, which is realised for a massless field
with $\tilde{\alpha}_{i}=0$, $i\neq l$, to the leading order we have
\begin{equation}
\langle j^{l}\rangle _{0}\approx \frac{2e\Gamma ((p+3)/2)}{\pi
^{(p+3)/2}L_{l}^{p+1}V_{q}}\sum_{n=1}^{\infty }\frac{\sin (n\tilde{\alpha}%
_{l})}{n^{p+2}}.  \label{jl0Llarge}
\end{equation}%
In this case, the leading term in the expansion of $V_{q}\langle
j^{l}\rangle _{0}/L_{l}$ coincides with the current density in $(p+2)$%
-dimensional space with a single compact dimension of the length $L_{l}$.
For $\omega _{0l}\neq 0$ and for large values of $L_{l}$ one has%
\begin{equation}
\langle j^{l}\rangle _{0}\approx \frac{2eV_{q}^{-1}\sin (\tilde{\alpha}%
_{l})\omega _{0l}^{p/2+1}}{(2\pi )^{p/2+1}L_{l}^{p/2}}e^{-L_{l}\omega _{0l}},
\label{jl0Llargeb}
\end{equation}%
and the current density is exponentially suppressed. In the opposite limit
of small values for $L_{l}$, to the leading order we get%
\begin{equation}
\langle j^{l}\rangle _{0}\approx \frac{2e\Gamma ((D+1)/2)}{\pi
^{(D+1)/2}L_{l}^{D}}\sum_{n=1}^{\infty }\frac{\sin (n\tilde{\alpha}_{l})}{%
n^{D}}.  \label{jrT0small}
\end{equation}%
The leading term does not depend on the mass and on the lengths of the other
compact dimensions and coincides with the current density for a massless
scalar field in the space with topology $R^{D-1}\times S^{1}$.

Now we turn to the investigation of the plate-induced contribution in the
current density. By using the expression for the corresponding part in the
Hadamard function from (\ref{G1jb}), we get the following expression%
\begin{equation}
\langle j^{l}\rangle _{j}^{(1)}=\frac{eC_{p}}{2^{p}V_{q}}\sum_{\mathbf{n}%
_{q}}k_{l}\int_{\omega _{\mathbf{n}_{q}}}^{\infty }dy\,(y^{2}-\omega _{%
\mathbf{n}_{q}}^{2})^{(p-1)/2}e^{-2yz_{j}}\frac{y\beta _{j}+1}{y\beta _{j}-1}%
,  \label{jlj(1)}
\end{equation}%
with the notations $z_{j}=|z-a_{j}|$ for the distance from the plate and%
\begin{equation}
C_{p}=\frac{\pi ^{-(p+1)/2}}{\Gamma ((p+1)/2)}.  \label{Cp}
\end{equation}%
Recall that, in order to have a stable vacuum state with $\langle \varphi
\rangle =0$, we have assumed that $1/\beta _{j}<\omega _{0}$. Under this
condition, the integrand in (\ref{jlj(1)}) is regular everywhere in the
integration range. The integral in (\ref{jlj(1)}) is evaluated in the
special cases of Dirichlet and Neumann boundary conditions with the result
\begin{equation}
\langle j^{l}\rangle _{j}^{(1)}=\mp \frac{2e/V_{q}}{(2\pi )^{p/2+1}}\sum_{%
\mathbf{n}_{q}}k_{l}\omega _{\mathbf{n}_{q}}^{p}f_{p/2}(2\omega _{\mathbf{n}%
_{q}}z_{j}),  \label{jlj(1)DN}
\end{equation}%
where the upper and lower signs correspond to Dirichlet and Neumann boundary
conditions, respectively. Note that, in the problem with a fermionic field,
obeying the bag boundary condition on the plate, the boundary-induced
contribution vanishes for a massless field \cite{Bell13b}.

Let us consider the behavior of the plate-induced contribution in asymptotic
regions of the parameters. At large distances from the plate, $z_{j}\gg
L_{i} $, one has $z_{j}\omega _{\mathbf{n}_{q}}\gg 1$. Assuming that $|%
\tilde{\alpha}_{i}|<\pi $, the dominant contribution in (\ref{jlj(1)}) comes
from the region near the lower limit of the integration and from the term
with $n_{i}=0$, $i=p+2,\ldots ,D$. To the leading order we find%
\begin{equation}
\langle j^{l}\rangle _{j}^{(1)}\approx \frac{e\tilde{\alpha}_{l}\omega
_{0}^{(p-1)/2}e^{-2\omega _{0}z_{j}}}{(4\pi
)^{(p+1)/2}V_{q}L_{l}z_{j}^{(p+1)/2}}\frac{\omega _{0}\beta _{j}+1}{\omega
_{0}\beta _{j}-1},  \label{jlj(1)far}
\end{equation}%
and the current density is exponentially small. Note that the suppression is
exponential for both massive and massless field.

For points close to the plate, $z_{j}\ll L_{i}$, in (\ref{jlj(1)}) the
contribution of the terms with large values of $|n_{i}|$ dominates and this
formula is not convenient for the asymptotic analysis and for numerical
evaluations. In the case $\beta _{j}\leqslant 0$, an alternative expression
is obtained by using the representation (\ref{G1alt3}) for the Hadamard
function. The first term in the right-hand side of this representation
corresponds to the geometry with uncompactified $l$th dimension and does not
contribute to the current density along that direction. In the geometry of a
single plate at $x^{p+1}=a_{j}$ the part in the Hadamard function induced by
the compactification is given by the first term in the figure braces of (\ref%
{G1alt3}). From this part, by making use of (\ref{jl1}), for the VEV of the $%
l$th component of the current density we get%
\begin{equation}
\langle j^{l}\rangle _{j}=\frac{2^{1-p/2}eL_{l}}{\pi ^{p/2+2}V_{q}}%
\sum_{n=1}^{\infty }\frac{\sin \left( n\tilde{\alpha}_{l}\right) }{%
(nL_{l})^{p+1}}\sum_{\mathbf{n}_{q-1}}\int_{0}^{\infty
}dy\,g(z_{j},y)g_{p/2+1}(nL_{l}\sqrt{y^{2}+\omega _{\mathbf{n}_{q-1}}^{2}}),
\label{jlalt}
\end{equation}%
where we have defined the function%
\begin{eqnarray}
g(z_{j},y) &=&g_{j}(z,z,y)=1+\frac{1}{2}\sum_{s=\pm 1}e^{2siyz_{j}}\frac{%
iy\beta _{j}-s}{iy\beta _{j}+s}  \notag \\
&=&1-\frac{(1-y^{2}\beta _{j}^{2})\cos (2yz_{j})+2y\beta _{j}\sin (2yz_{j})}{%
1+y^{2}\beta _{j}^{2}}.  \label{gzj}
\end{eqnarray}%
The part with the first term in the right-side of (\ref{gzj}) corresponds to
the current density in the boundary-free geometry. In this part the
integration over $y$ is done with the help of the formula%
\begin{equation}
\int_{0}^{\infty }dy\,g_{\frac{p}{2}+1}(nL_{l}\sqrt{y^{2}+b^{2}})=\sqrt{\pi
/2}(nL_{l})^{-1}g_{\frac{p+3}{2}}(nL_{l}b),  \label{IntRel}
\end{equation}%
and one gets the expression (\ref{jl0b}).

Extracing the boundary-free part, for the plate-induced contribution from (%
\ref{jlalt}) we find%
\begin{equation}
\langle j^{l}\rangle _{j}^{(1)}=\frac{2^{-p/2}eL_{l}}{\pi ^{p/2+2}V_{q}}%
\sum_{n=1}^{\infty }\frac{\sin \left( n\tilde{\alpha}_{l}\right) }{%
(nL_{l})^{p+1}}\sum_{\mathbf{n}_{q-1}}\int_{0}^{\infty }dy\,g_{\frac{p}{2}%
+1}(nL_{l}\sqrt{y^{2}+\omega _{\mathbf{n}_{q-1}}^{2}})\sum_{s=\pm
1}e^{2siyz_{j}}\frac{iy\beta _{j}-s}{iy\beta _{j}+s}.  \label{jl1alt}
\end{equation}%
In the case of single compact dimension one has $q=1$, $p=D-2$, and the
corresponding formula for the plate-induced contribution in the current
density is obtained from (\ref{jl1alt}) omitting the summation over $\mathbf{%
n}_{q-1}$ and putting $\omega _{\mathbf{n}_{q-1}}=m$.

An important issue in quantum field theory with boundaries is the appearance
of surface divergences in the VEVs of local physical observables. Examples
of the latter are the VEVs of the field squared and of the energy density.
These divergences are a consequence of the oversimplification of a model
where the physical interactions are replaced by the imposition of boundary
conditions for all modes of a fluctuating quantum field. Of course, this is
an idealization, as real physical systems cannot constrain all the modes
(for a discussion of surface divergences and their physical interpretation
see \cite{Most97,Deut79} and references therein). The appearance of
divergences in the VEVs of physical quantities indicates that a more
realistic physical model should be employed for their evaluation on the
boundaries. An important feature, which directly follows from the
representation (\ref{jl1alt}), is that the VEV of the current density is
finite on the plate. This is in sharp contrast with the behavior of the VEVs
for the field squared and energy-momentum tensor. The finiteness of the
current density on the boundary may be understood from general arguments.
The divergences in local physical observables are determined by the local
bulk and boundary geometries. If we consider the model with the topology $%
R^{p+2}\times T^{q-1}$ with the $l$th dimension having the topology $R^{1}$,
then in this model the $l$th component of the current density vanishes by
the symmetry. The compactification of the $l$th dimension to $S^{1}$ does
not change both the bulk end boundary local geometries and, hence, does not
add new divergences to the VEVs compared with the model on $R^{p+2}\times
T^{q-1}$.

In deriving (\ref{jl1alt}) we have assumed that $\beta _{j}\leqslant 0$. In
the case $\beta _{j}>0$ the contribution of the bound state should be added
to (\ref{jl1alt}). For $1/\beta _{j}<\omega _{0l}$, this contribution is
obtained from the corresponding part in the Hadamard function, given by (\ref%
{Gjbound}), and has the form%
\begin{equation}
\langle j_{l}\rangle _{bj}^{(1)}=-\frac{2^{2-p/2}eL_{l}e^{-2z_{j}/\beta _{j}}%
}{\pi ^{p/2+1}V_{q}\beta _{j}}\sum_{n=1}^{\infty }\frac{\sin \left( n\tilde{%
\alpha}_{l}\right) }{(nL_{l})^{p+1}}\sum_{\mathbf{n}_{q-1}}g_{\frac{p}{2}%
+1}(nL_{l}\sqrt{\omega _{\mathbf{n}_{q-1}}^{2}-1/\beta _{j}^{2}}).
\label{jl1bound}
\end{equation}%
In what follows for simplicity we shall consider the case $\beta
_{j}\leqslant 0$. Recall that, the representation (\ref{jlj(1)}) is valid
for all values of $\beta _{j}$ from the range of the vacuum stability.

For Dirichlet and Neumann boundary conditions, after the evaluation of the
integral in (\ref{jl1alt}) by using the formula%
\begin{equation}
\int_{0}^{\infty }dy\,\cos (2yz_{j})g_{\frac{p}{2}+1}(nL_{l}\sqrt{y^{2}+b^{2}%
})=\sqrt{\frac{\pi }{2}}(nL_{l})^{p+2}\frac{g_{\frac{p+3}{2}}(b\sqrt{%
4z_{j}^{2}+n^{2}L_{l}^{2}})}{(4z_{j}^{2}+n^{2}L_{l}^{2})^{(p+3)/2}},
\label{IntRel2}
\end{equation}%
one gets%
\begin{equation}
\langle j^{l}\rangle _{j}^{(1)}=\mp \frac{4eL_{l}^{2}/V_{q}}{(2\pi
)^{(p+3)/2}}\sum_{n=1}^{\infty }\frac{n\sin \left( n\tilde{\alpha}%
_{l}\right) }{(4z_{j}^{2}+n^{2}L_{l}^{2})^{(p+3)/2}}\sum_{\mathbf{n}%
_{q-1}}g_{\frac{p+3}{2}}(\omega _{\mathbf{n}_{q-1}}\sqrt{%
4z_{j}^{2}+n^{2}L_{l}^{2}}),  \label{jl1altDN}
\end{equation}%
where the upper and lower signs correspond to Dirichlet and Neumann
conditions, respectively. For a single compact dimension with the length $L$
and with the phase $\tilde{\alpha}$ in the periodicity condition for a
massless field this gives%
\begin{equation}
\langle j^{l}\rangle _{j}^{(1)}=\mp \frac{2\Gamma ((D+1)/2)e}{\pi
^{(D+1)/2}L^{D}}\sum_{n=1}^{\infty }\frac{n\sin \left( n\tilde{\alpha}%
\right) }{(n^{2}+4z_{j}^{2}/L^{2})^{(D+1)/2}}.  \label{jl1DNm0}
\end{equation}%
Now, combining the expressions (\ref{jl0b}) and (\ref{jl1altDN}), we see
that in the case of Dirichlet boundary condition the boundary-free and
plate-induced parts of the current density cancel each other for $z_{j}=0$
and, hence, the total current vanishes on the plate. For Neumann condition
the current density on the plate is given by
\begin{equation}
\langle j^{l}\rangle _{j,z=a_{j}}=2\langle j^{l}\rangle _{0}=\frac{%
8eL_{l}/V_{q}}{(2\pi )^{(p+3)/2}}\sum_{n=1}^{\infty }\frac{\sin \left( n%
\tilde{\alpha}_{l}\right) }{(nL_{l})^{p+2}}\sum_{\mathbf{n}_{q-1}}g_{\frac{%
p+3}{2}}(nL_{l}\omega _{\mathbf{n}_{q-1}}).  \label{jlNpl}
\end{equation}%
Note that the normal derivative of the current density on the plate vanishes
for both Dirichlet and Neumann boundary conditions: $(\partial _{z}\langle
j^{l}\rangle _{j})_{z=a_{j}}=0$. This is not the case for general Robin
condition.

Let us consider the behavior of the plate-induced contribution in the
current density in the limit $L_{i}\ll L_{l}$. In this investigation it is
more convenient to use the representation (\ref{jl1alt}). For $%
\sum\nolimits_{i=p+2,\neq l}^{D}\tilde{\alpha}_{i}^{2}\neq 0$, the dominant
contribution in the integral of (\ref{jl1alt}) comes from the region near
the lower limit of the integration and from the term $n=1$, $n_{i}=0$, $%
i=p+2,\ldots ,D$, in the summation. The argument of the function $%
g_{p/2+1}(x)$ in the integrand is large and we can use the asymptotic
expression $g_{\nu }(x)\approx \sqrt{\pi /2}x^{\nu -1/2}e^{-x}$. After some
intermediate calculations, for the leading term we get%
\begin{equation}
\langle j^{l}\rangle _{j}^{(1)}\approx \frac{2e(1-2\delta _{0\beta _{j}})}{%
(2\pi )^{p/2+1}V_{q}L_{l}^{p/2}}\frac{\omega _{0l}^{p/2+1}\sin \tilde{\alpha}%
_{l}}{e^{L_{l}\omega _{0l}(1+2z_{j}^{2}/L_{l}{}^{2})}}.  \label{jl1SmalLr}
\end{equation}%
Here, we have additionally assumed that $L_{i}\ll |\beta _{j}|$ for $\beta
_{j}\neq 0$. For $\tilde{\alpha}_{i}=0$, $i=p+2,\ldots ,D$, $i\neq l$, the
dominant contribution in (\ref{jl1alt}) comes from the term $n_{i}=0$, $%
i=p+2,\ldots ,D$, with the leading term
\begin{eqnarray}
\frac{V_{q}}{L_{l}}\langle j^{l}\rangle _{j}^{(1)} &\approx &\langle
j^{l}\rangle _{j,R^{p+1}\times S^{1}}^{(1)}=\frac{4e}{(2\pi )^{p/2+2}}%
\sum_{n=1}^{\infty }\frac{\sin \left( n\tilde{\alpha}_{l}\right) }{%
(nL_{l})^{p+1}}\int_{0}^{\infty }dy\,  \notag \\
&&\times g_{\frac{p}{2}+1}(nL_{l}\sqrt{y^{2}+m^{2}})\sum_{s=\pm
1}e^{2siyz_{j}}\frac{iy\beta _{j}-s}{iy\beta _{j}+s}.  \label{jl1SmalLrb}
\end{eqnarray}%
Here, $\langle j^{l}\rangle _{j,R^{p+1}\times S^{1}}^{(1)}$ is the
plate-induced contribution in the current density for $(p+2)$-dimensional
space with topology $R^{p+1}\times S^{1}$ (see (\ref{jl1alt}) for the case $%
q=1$ and, hence, $\omega _{\mathbf{n}_{q-1}}=m$).

If the length of the $i$th compact dimension is large, $i\neq l$, the
dominant contribution to the sum over $n_{i}$ comes from large values of $%
|n_{i}|$ and in (\ref{jl1alt}) we can replace the summation over $n_{i}$ by
the integration in accordance with%
\begin{equation}
\sum_{n_{i}=-\infty }^{\infty }f(\vert k_{i}\vert )\rightarrow \frac{L_{i}}{%
\pi }\int_{0}^{\infty }dx\,\,f(x).  \label{Repl}
\end{equation}%
The integral over $x$ is evaluated by using the formula (\ref{IntRel}). As a
result, from (\ref{jl1alt}), to the leading order, we obtain the current
density along the $l$th compact dimension for the spatial topology $%
R^{p+2}\times T^{q-1}$ with the lengths of the compact dimensions $%
(L_{p+2},\ldots ,L_{i-1},L_{i+1},\ldots ,L_{D})$.

Now let us consider the limiting case when $L_{l}$ is large compared with
the other length scales in the problem, $L_{l}\gg L_{i},z_{j}$, $i\neq l$.
The dominant contribution in (\ref{jl1alt}) comes from the term $n_{i}=0$, $%
i\neq l$. For $\omega _{0l}\neq 0$ we find%
\begin{equation}
\langle j^{l}\rangle _{j}^{(1)}\approx \frac{2e\left( 2\delta _{\beta
_{j},\infty }-1\right) }{(2\pi )^{p/2+1}V_{q}}\frac{\sin \tilde{\alpha}_{l}}{%
L_{l}{}^{p/2}}\omega _{0l}^{p/2+1}e^{-L_{l}\omega _{0l}},  \label{jlLlarge}
\end{equation}%
where, for non-Neumann boundary conditions ($\beta _{j}\neq \infty $), we
have assumed that $\beta _{j}\omega _{0l}\ll \left( L_{l}\omega _{0l}\right)
^{1/2}$. For $\omega _{0l}=0$ the leading term is given by the expression%
\begin{equation}
\langle j^{l}\rangle _{j}^{(1)}\approx \frac{2e\left( 2\delta _{\beta
_{j},\infty }-1\right) }{\pi ^{(p+3)/2}V_{q}L_{l}^{p+1}}\Gamma
((p+3)/2)\sum_{n=1}^{\infty }\frac{\sin \left( n\tilde{\alpha}_{l}\right) }{%
n^{p+2}}.  \label{jlLlargeb}
\end{equation}%
Comparing with the corresponding asymptotics (\ref{jl0Llarge}) and (\ref%
{jl0Llargeb}), we see that for non-Neumann boundary conditions, in the both
cases $\omega _{0l}\neq 0$ and $\omega _{0l}=0$, the leading terms in the
boundary-induced and boundary-free parts of the current density cancel each
other.

An equivalent representation for the plate-induced current density is
obtained from (\ref{jl1alt}) rotating the integration contour in the complex
plane $y$ by the angle $\pi /2$ for the term with $s=1$ and by the angle $%
-\pi /2$ for the term with $s=-1$. The integrals over the intervals $(0,\pm
i\omega _{\mathbf{n}_{q-1}})$ are cancelled and we find%
\begin{eqnarray}
\langle j^{l}\rangle _{j}^{(1)} &=&\frac{2^{-p/2}eL_{l}}{\pi ^{p/2+1}V_{q}}%
\sum_{n=1}^{\infty }\frac{\sin \left( n\tilde{\alpha}_{l}\right) }{%
(nL_{l})^{p+1}}\sum_{\mathbf{n}_{q-1}}\int_{\omega _{\mathbf{n}%
_{q-1}}}^{\infty }dy\,  \notag \\
&&\times e^{-2yz_{j}}\frac{y\beta _{j}+1}{y\beta _{j}-1}w_{p/2+1}(nL_{l}%
\sqrt{y^{2}-\omega _{\mathbf{n}_{q-1}}^{2}}),  \label{jl1alt3}
\end{eqnarray}%
where%
\begin{equation}
w_{\nu }(x)=x^{\nu }J_{\nu }(x),  \label{w}
\end{equation}%
and $J_{\nu }(x)$ is the Bessel function. The equivalence of the
representations (\ref{jlj(1)}) and (\ref{jl1alt3}) can also be directly seen
by applying to the series over $n_{l}$ in (\ref{jlj(1)}) the relation%
\begin{equation}
\sum_{n_{l}=-\infty }^{+\infty }k_{l}g(|k_{l}|)=\frac{2L_{l}}{\pi }%
\sum_{n=1}^{\infty }\sin (n\tilde{\alpha}_{l})\int_{0}^{\infty }dx\,x\sin
(nL_{l}x)g(x).  \label{SumForm3}
\end{equation}%
The latter is a direct consequence of the Poisson's resummation formula.
After using (\ref{SumForm3}) in (\ref{jlj(1)}), we introduce a new
integration variable $u=\sqrt{y^{2}-x^{2}-\omega _{\mathbf{n}_{q-1}}^{2}}$
and then pass to polar coordinates in the $(u,x)$-plane. The integration
over the polar angle is expressed in terms of the Bessel function and the
representation (\ref{jl1alt3}) is obtained.

Another expression is obtained by applying to the series over $n_{l}$ in (%
\ref{jlj(1)}) the summation formula (\ref{AbelPlan1}). For the series in (%
\ref{jlj(1)}) one has $g(u)=u$ and the first integral vanishes. As a result,
the plate-induced part in the VEV of the current density is presented as%
\begin{eqnarray}
\langle j^{l}\rangle _{j}^{(1)} &=&-\frac{eC_{p}L_{l}\sin \tilde{\alpha}_{l}%
}{2^{p}\pi V_{q}}\sum_{\mathbf{n}_{q-1}}\int_{0}^{\infty }dx\,\frac{x}{\cosh
(L_{l}\sqrt[.]{x^{2}+\omega _{\mathbf{n}_{q-1}}^{2}})-\cos \tilde{\alpha}_{l}%
}  \notag \\
&&\times \int_{0}^{x}dy\frac{(1-y^{2}\beta _{j}^{2})\cos \left(
2yz_{j}\right) +2y\beta _{j}\sin \left( 2yz_{j}\right) }{(1+y^{2}\beta
_{j}^{2})\left( x^{2}-y^{2}\right) ^{(1-p)/2}}.  \label{jlj(1)Alt}
\end{eqnarray}%
For Dirichlet and Neumann boundary conditions we obtain%
\begin{equation}
\langle j^{l}\rangle _{j}^{(1)}=\mp \frac{2eL_{l}\sin \tilde{\alpha}_{l}}{%
(4\pi )^{p/2+1}V_{q}z_{j}^{p/2}}\sum_{\mathbf{n}_{q-1}}\int_{0}^{\infty }dx\,%
\frac{x^{p/2+1}\,J_{p/2}(2xz_{j})}{\cosh (L_{l}\sqrt[.]{x^{2}+\omega _{%
\mathbf{n}_{q-1}}^{2}})-\cos \tilde{\alpha}_{l}}.  \label{jlj(1)AltDN}
\end{equation}

\begin{figure}[tbph]
\begin{center}
\begin{tabular}{cc}
\epsfig{figure=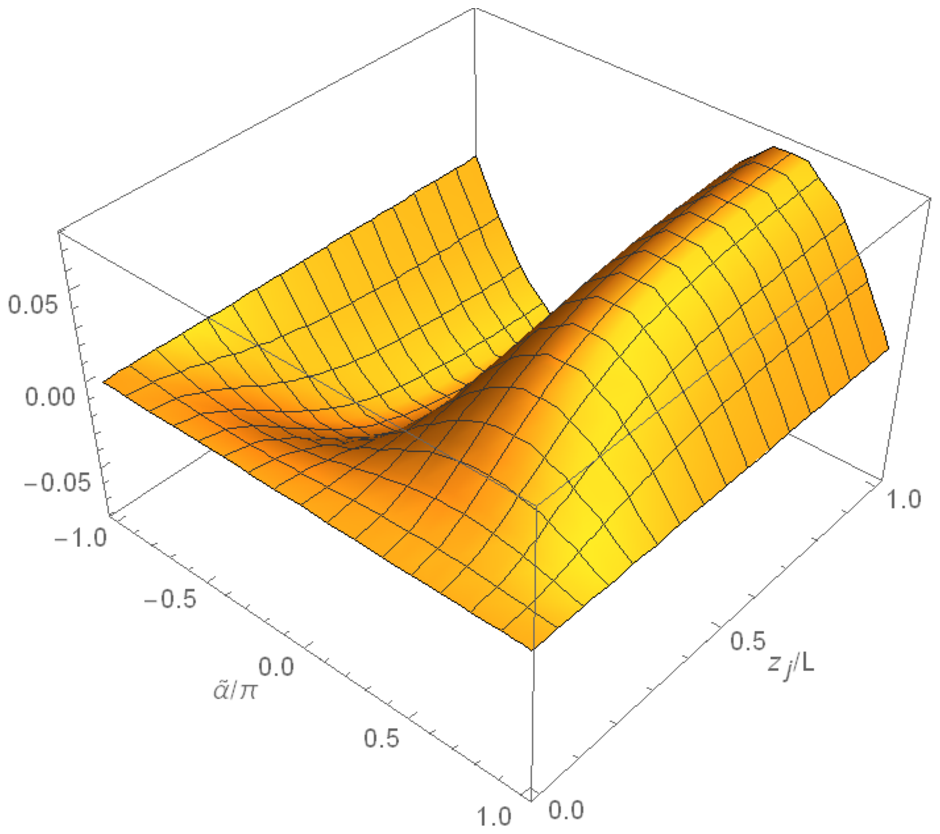,width=7.cm,height=6.cm} & \quad %
\epsfig{figure=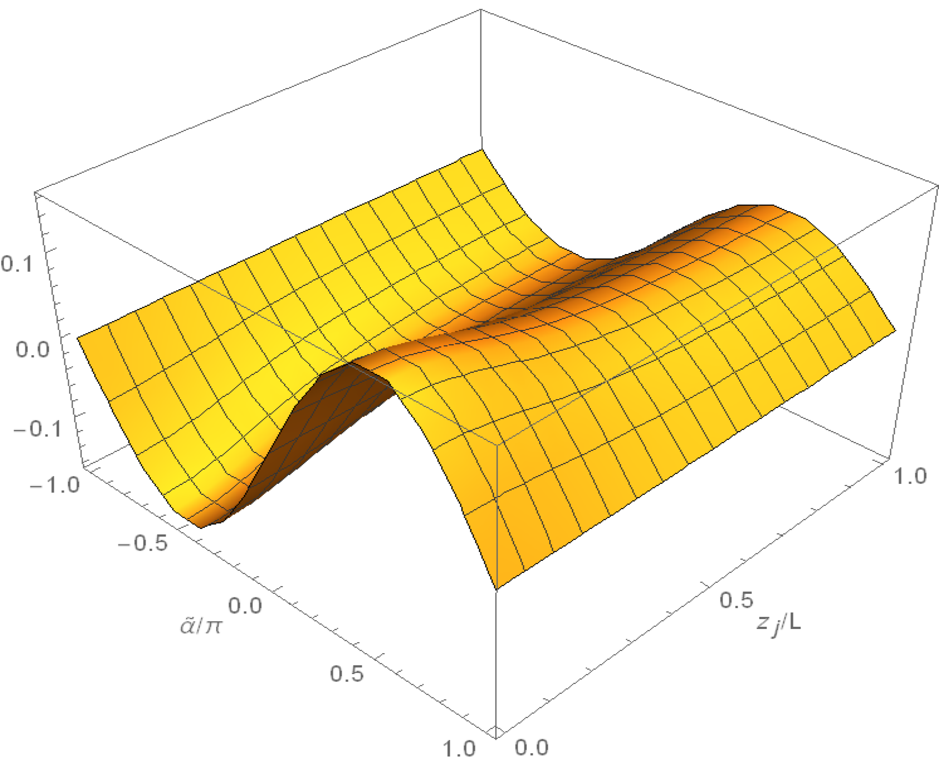,width=7.cm,height=6.cm}%
\end{tabular}%
\end{center}
\caption{The total current density, $L^{D}\langle j^{l}\rangle _{j}/e$, in
the topology $R^{3}\times S^{1}$ for a $D=4$ massless scalar field with
Dirichlet (left panel) and Neumann (right panel) boundary conditions in the
geometry of a single plate, as a function of the phase in the
quasiperiodicity boundary condition and of the distance from the plate.}
\label{fig1}
\end{figure}

In figure \ref{fig1}, for the simplest Kaluza-Klein model with a single
compact dimension of the length $L$ and with the phase $\tilde{\alpha}$ ($D=4
$), we have plotted the total current density, $L^{D}\langle j^{l}\rangle
_{j}/e$, for a massless scalar field in the geometry of a single plate as a
function of the distance from the plate and of the phase $\tilde{\alpha}$.
The left/right panel correspond to Dirichlet/Neumann boundary conditions. As
has been already noticed before, in the Dirichlet case the total current
density vanishes on the plate.

\begin{figure}[tbph]
\begin{center}
\begin{tabular}{cc}
\epsfig{figure=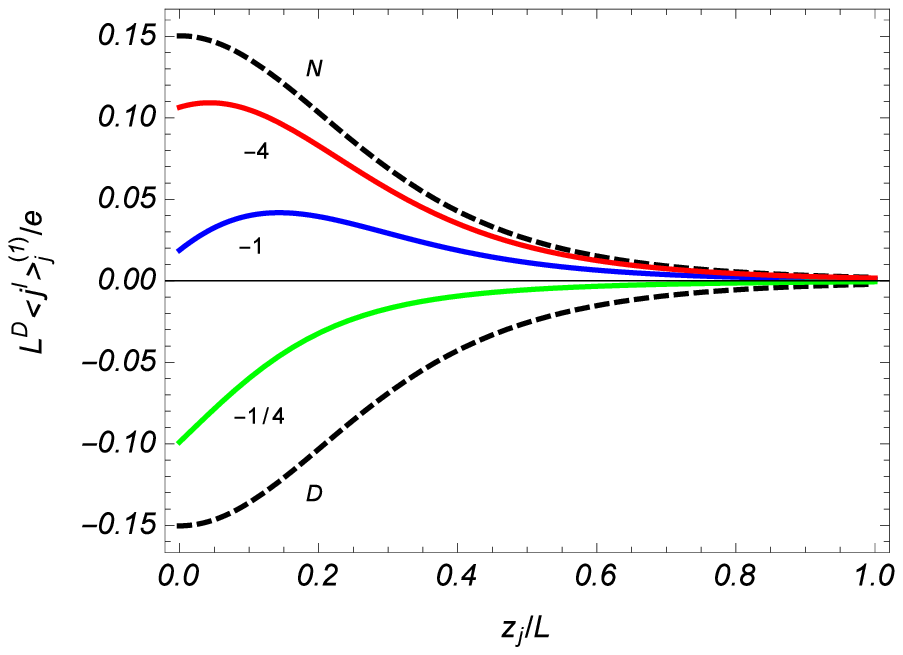,width=7.cm,height=6.cm} & \quad %
\epsfig{figure=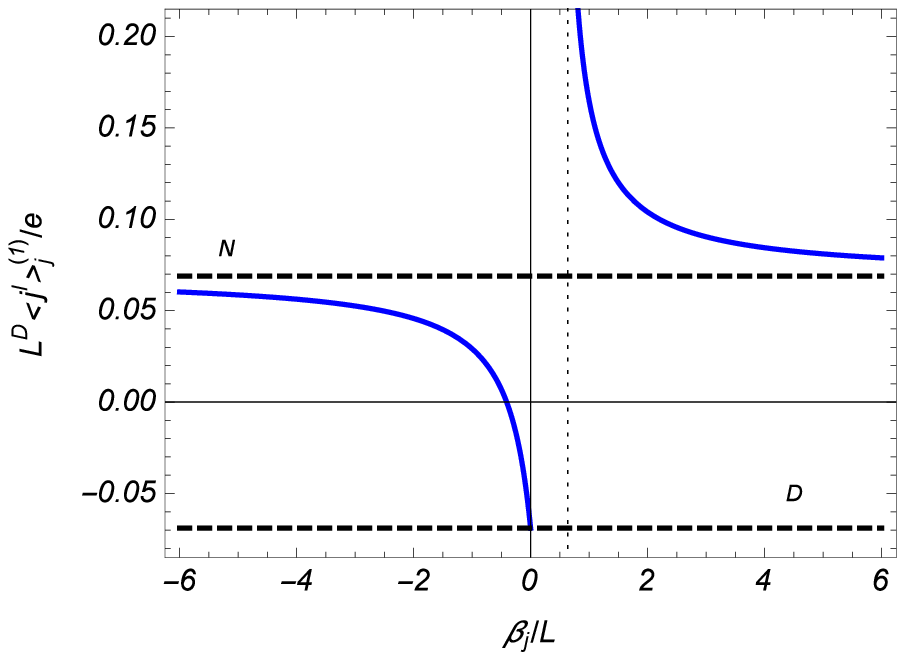,width=7.cm,height=6.cm}%
\end{tabular}%
\end{center}
\caption{The plate-induced contribution to the current density for the model
corresponding to figure \protect\ref{fig1} as a function of the distance
from the plate (left panel) for different values of the ratio $\protect\beta %
_{j}/L$ (numbers near the curves) and as a function of $\protect\beta _{j}/L$
(right panel) for $z_{j}/L=0.3$. The dashed curves correspond to Dirichlet
and Neumann boundary conditions and the graphs are plotted for $\tilde{%
\protect\alpha}=\protect\pi /2$.}
\label{fig2}
\end{figure}

For the same model, figure \ref{fig2} presents the plate-induced
contribution to the current density as a function of the distance from the
plate for various values of the coefficients in the Robin boundary condition
(left panel) and as a function of the ratio $\beta _{j}/L$ (right panel).
The numbers near the curves on the right panel correspond to the value of $%
\beta _{j}/L$. The left panel is plotted for the fixed value of the relative
distance from the plate $z_{j}/L=0.3$. On both panels, the dashed curves are
plotted for Dirichlet and Neumann boundary conditions. For the phase in the
quasiperiodicity condition we have taken $\tilde{\alpha}=\pi /2$. On the
right panel, for the values of $\beta _{j}/L$ between the ordinate axis and
the vertical dotted line ($\beta _{j}/L=1/\tilde{\alpha}$) the vacuum is
unstable.

\section{Current density between two plates}

\label{sec:2Plates}

Now we turn to the geometry of two plates. In the region $a_{1}\leqslant
x^{p+1}\leqslant a_{2}$, by using the formula (\ref{G(1)}) for the Hadamard
function, the VEV of the current density is decomposed as
\begin{equation}
\langle j^{l}\rangle =\langle j^{l}\rangle _{j}+\frac{eC_{p}}{2^{p-1}V_{q}}%
\sum_{\mathbf{n}_{q}}k_{l}\int_{\omega _{\mathbf{n}_{q}}}^{\infty }dy\frac{%
(y^{2}-\omega _{\mathbf{n}_{q}}^{2})^{(p-1)/2}g(z_{j},iy)}{%
c_{1}(ay)c_{2}(ay)e^{2ay}-1}.  \label{jl2}
\end{equation}%
Here, the second term in the right-hand side is induced by the plate at $%
x^{p+1}=a_{j^{\prime }}$, $j^{\prime }\neq j$.

Extracting from the second term in the right-hand side of (\ref{jl2}) the
part induced by the second plate when the first one is absent, the current
density is written in a more symmetric form:%
\begin{equation}
\langle j^{l}\rangle =\langle j^{l}\rangle _{0}+\sum_{j=1,2}\langle
j^{l}\rangle _{j}^{(1)}+\Delta \langle j^{l}\rangle ,  \label{jldec}
\end{equation}%
where the interference part is given by the expression%
\begin{equation}
\Delta \langle j^{l}\rangle =\frac{eC_{p}}{2^{p}V_{q}}\sum_{\mathbf{n}%
_{q}}k_{l}\int_{\omega _{\mathbf{n}_{q}}}^{\infty }dy\,(y^{2}-\omega _{%
\mathbf{n}_{q}}^{2})^{\frac{p-1}{2}}\frac{2+%
\sum_{j=1,2}e^{-2yz_{j}}/c_{j}(ay)}{c_{1}(ay)c_{2}(ay)e^{2ay}-1}.
\label{jlint1}
\end{equation}%
By taking into account the expression for the current density in the
geometry of a single plate, for the total current density we can also write%
\begin{eqnarray}
\langle j^{l}\rangle &=&\langle j^{l}\rangle _{0}+\frac{eC_{p}}{2^{p}V_{q}}%
\sum_{\mathbf{n}_{q}}k_{l}\int_{\omega _{\mathbf{n}_{q}}}^{\infty
}dy\,(y^{2}-\omega _{\mathbf{n}_{q}}^{2})^{\frac{p-1}{2}}  \notag \\
&&\times \frac{2+\sum_{j=1,2}c_{j}(ay)e^{2yz_{j}}}{%
c_{1}(ay)c_{2}(ay)e^{2ay}-1}.  \label{jltot}
\end{eqnarray}

For special cases of Dirichlet and Neumann boundary conditions on both
plates the general formula is simplified to%
\begin{equation}
\langle j^{l}\rangle =\langle j^{l}\rangle _{0}+\frac{eC_{p}}{2^{p}V_{q}}%
\sum_{\mathbf{n}_{q}}k_{l}\int_{\omega _{\mathbf{n}_{q}}}^{\infty
}dy\,(y^{2}-\omega _{\mathbf{n}_{q}}^{2})^{\frac{p-1}{2}}\frac{2\mp
\sum_{j=1,2}e^{2yz_{j}}}{e^{2ay}-1},  \label{jl2DN}
\end{equation}%
where, as before, the upper and lower signs correspond to Dirichlet and
Neumann boundary conditions, respectively. In particular, for Dirichlet
boundary condition the part induced by the second plate vanishes on the
first plate. Note that in the system of two fields with Dirichlet and
Neumann conditions the distribution of the total current density in the
region between the plates is uniform and the current density vanishes in the
regions $z<a_{1}$ and $z>a_{2}$. Another form for (\ref{jl2DN}) is obtained
by making use of the expansion%
\begin{equation}
\frac{1}{e^{2ay}-1}=\sum_{n=1}^{\infty }e^{-2nay},  \label{Exp}
\end{equation}%
After the integration over $y$ we get
\begin{equation}
\langle j^{l}\rangle =\langle j^{l}\rangle _{0}+\frac{2e/V_{q}}{(2\pi
)^{p/2+1}}\sum_{n=1}^{\infty }\sum_{\mathbf{n}_{q}}k_{l}\omega _{\mathbf{n}%
_{q}}^{p}[2f_{\frac{p}{2}}(2na\omega _{\mathbf{n}_{q}})\mp \sum_{j=1,2}f_{%
\frac{p}{2}}(2(na-z_{j})\omega _{\mathbf{n}_{q}})].  \label{jl2DN2}
\end{equation}%
A similar representation for the interference part $\Delta \langle
j^{l}\rangle $ is obtained from (\ref{jl2DN2}) by the replacement $%
z_{j}\rightarrow -z_{j}$. For Dirichlet boundary condition, on the plates, $%
z=a_{j}$, one has%
\begin{equation}
\Delta \langle j^{l}\rangle _{z=a_{j}}=\frac{2e/V_{q}}{(2\pi )^{p/2+1}}\sum_{%
\mathbf{n}_{q}}k_{l}\omega _{\mathbf{n}_{q}}^{p}f_{\frac{p}{2}}(2a\omega _{%
\mathbf{n}_{q}}).  \label{jintDpl}
\end{equation}%
Combining this result with the formulas for single plates, we see that in
the case of Dirichlet boundary condition the total current vanishes on the
plates: $\langle j^{l}\rangle _{z=a_{j}}=0$.

An equivalent representation for the current density in the region between
the plates and for Robin conditions is obtained by using the representation (%
\ref{G1alt3}) for the corresponding Hadamard function:%
\begin{eqnarray}
\langle j^{l}\rangle  &=&\langle j^{l}\rangle _{j}+\frac{2^{1-p/2}eL_{l}}{%
\pi ^{p/2+1}V_{q}}\sum_{n=1}^{\infty }\frac{\sin \left( n\tilde{\alpha}%
_{l}\right) }{(nL_{l})^{p+1}}\sum_{\mathbf{n}_{q-1}}\int_{\omega _{\mathbf{n}%
_{q-1}}}^{\infty }dy  \notag \\
&&\times \frac{w_{p/2+1}(nL_{l}\sqrt{y^{2}-\omega _{\mathbf{n}_{q-1}}^{2}})}{%
c_{1}(ay)c_{2}(ay)e^{2ay}-1}g(z_{j},iy).  \label{jlalt2}
\end{eqnarray}%
Combining the expressions (\ref{jl1alt3}) and (\ref{jlalt2}), for the total
current density we find%
\begin{eqnarray}
\langle j^{l}\rangle  &=&\langle j^{l}\rangle _{0}+\frac{2^{-p/2}eL_{l}}{\pi
^{p/2+1}V_{q}}\sum_{n=1}^{\infty }\frac{\sin \left( n\tilde{\alpha}%
_{l}\right) }{(nL_{l})^{p+1}}\sum_{\mathbf{n}_{q-1}}\int_{\omega _{\mathbf{n}%
_{q-1}}}^{\infty }dy\,  \notag \\
&&\times \frac{2+\sum_{j=1,2}e^{2yz_{j}}c_{j}(ay)}{%
c_{1}(ay)c_{2}(ay)e^{2ay}-1}w_{p/2+1}(nL_{l}\sqrt{y^{2}-\omega _{\mathbf{n}%
_{q-1}}^{2}}).  \label{jltotalt}
\end{eqnarray}%
Now, by taking into account the expression (\ref{jl1alt3}) for the single
plate induced part, from (\ref{jlalt2}) for the interference part we get%
\begin{eqnarray}
\Delta \langle j^{l}\rangle  &=&\frac{2^{-p/2}eL_{l}}{\pi ^{p/2+1}V_{q}}%
\sum_{n=1}^{\infty }\frac{\sin \left( n\tilde{\alpha}_{l}\right) }{%
(nL_{l})^{p+1}}\sum_{\mathbf{n}_{q-1}}\int_{\omega _{\mathbf{n}%
_{q-1}}}^{\infty }dy\,  \notag \\
&&\times \frac{2+\sum_{j=1,2}e^{-2yz_{j}}/c_{j}(ay)}{%
c_{1}(ay)c_{2}(ay)e^{2ay}-1}w_{p/2+1}(nL_{l}\sqrt{y^{2}-\omega _{\mathbf{n}%
_{q-1}}^{2}}).  \label{jlint2}
\end{eqnarray}%
The equivalence of the representations (\ref{jltot}) and (\ref{jlalt2}) can
be seen directly by using the formula (\ref{SumForm3}) in a way similar to
that for the geometry of a single plate.

For Dirichlet and Neumann conditions, after using the expansion (\ref{Exp}),
the integral over $y$ in (\ref{jltotalt}) is expressed in terms of the
MacDonald function and one gets the representation%
\begin{eqnarray}
\langle j^{l}\rangle &=&\frac{2^{(1-p)/2}eL_{l}^{2}}{\pi ^{(p+3)/2}V_{q}}%
\sum_{n=1}^{\infty }n\sin \left( n\tilde{\alpha}_{l}\right) \sum_{\mathbf{n}%
_{q-1}}\omega _{\mathbf{n}_{q-1}}^{p+3}  \notag \\
&&\times \sum_{r=-\infty }^{\infty }\left\{ f_{\frac{p+3}{2}}(\omega _{%
\mathbf{n}_{q-1}}\sqrt{4(ra)^{2}+n^{2}L_{l}^{2}})\right.  \notag \\
&&\left. \mp f_{\frac{p+3}{2}}(\omega _{\mathbf{n}_{q-1}}\sqrt{%
4(ra-z+a_{1})^{2}+n^{2}L_{l}^{2}})\right\} ,  \label{jlDN}
\end{eqnarray}%
where we have taken into account the expression (\ref{jl0b}) for the current
density in the boundary-free geometry. In the model with a single compact
dimension with the length $L$ and for a massless field, from (\ref{jlDN}) we
find%
\begin{eqnarray}
\langle j^{l}\rangle &=&\frac{2\Gamma ((D+1)/2)e}{\pi ^{(D+1)/2}L^{D}}%
\sum_{n=1}^{\infty }\sum_{r=-\infty }^{\infty }n\sin \left( n\tilde{\alpha}%
\right)  \notag \\
&&\times \left\{ \left[ 4(ra/L)^{2}+n^{2}\right] ^{-\frac{D+1}{2}}\mp \left[
4(ra-z+a_{1})^{2}/L^{2}+n^{2}\right] ^{-\frac{D+1}{2}}\right\} .
\label{jlDNaltm0}
\end{eqnarray}%
In the case of Dirichlet boundary condition on the left plate, $%
x^{p+1}=a_{1} $, and Neumann boundary condition on the right one, $%
x^{p+1}=a_{2}$, the corresponding formulas are obtained from (\ref{jlDN})
and (\ref{jlDNaltm0}) with the upper sign, adding the factor $(-1)^{r}$ in
the summation over $r$. The corresponding current density vanishes on the
left plate. From (\ref{jlDN}) we can also see that the normal derivative of
the current density vanishes on the plates for both Dirichlet and Neumann
boundary conditions.

In the limit $a\ll L_{i}$, $i\neq l$, the dominant contribution to the
series over $\mathbf{n}_{q-1}$ in (\ref{jlint2}) comes from large values of $%
|n_{i}|$, $i\neq l$, and we can replace the summation by the integration in
accordance with%
\begin{equation}
\sum_{\mathbf{n}_{q-1}}f(\omega _{\mathbf{n}_{q-1}})\rightarrow \frac{%
2\left( 4\pi \right) ^{(1-q)/2}V_{q}}{\Gamma ((q-1)/2)L_{l}}\int_{0}^{\infty
}du\,u^{q-2}\,f(\sqrt{u^{2}+m^{2}}).  \label{SumRepl}
\end{equation}%
Changing the integration variable $y$ to $x=\sqrt{y^{2}-u^{2}}$, we
introduce polar coordinates in the $(u,x)$-plane. After the integration over
the polar angle, we get%
\begin{equation}
\Delta \langle j^{l}\rangle \approx \Delta \langle j^{l}\rangle
_{R^{D}\times S^{1}},  \label{jlsmalla}
\end{equation}%
where $\Delta \langle j^{l}\rangle _{R^{D}\times S^{1}}$ is the
corresponding quantity in the geometry of a single compact dimension with
the length $L_{l}$. The expression for $\Delta \langle j^{l}\rangle
_{R^{D}\times S^{1}}$ is obtained from (\ref{jlint2}) taking $p=D-2$, $%
V_{q}=L_{l}$, $\omega _{\mathbf{n}_{q-1}}=m$, and omitting the summation
over $\mathbf{n}_{q-1}$. If, in addition, $am\ll 1$, one finds%
\begin{equation}
\Delta \langle j^{l}\rangle \approx \frac{2e}{\left( 2\pi \right) ^{D/2}a}%
\sum_{n=1}^{\infty }\frac{\sin \left( n\tilde{\alpha}_{l}\right) }{%
(nL_{l})^{D-1}}\int_{0}^{\infty }dy\,\frac{2+%
\sum_{j=1,2}e^{-2yz_{j}/a}/c_{j}(y)}{c_{1}(y)c_{2}(y)e^{2y}-1}%
w_{D/2}(nL_{l}y/a).  \label{jlsmalla2}
\end{equation}

Now let us also assume that $a\ll L_{i},m^{-1}$, for all $i=p+2,\ldots ,D$.
This means that the separation between the plates is smaller than all other
length scales in the problem. In order to estimate the integral in (\ref%
{jlsmalla2}), we note that for a fixed $b$ and for $\lambda \rightarrow
+\infty $, the dominant contribution to the integral $\int_{0}^{\infty
}dy\,f(y)e^{-by}w_{D/2}(\lambda y)$ comes from the region with $y\lesssim a/L
$. By taking into account that%
\begin{equation}
\int_{0}^{\infty }dy\,e^{-by}w_{D/2}(\lambda y)=\frac{2^{D/2}\lambda
^{D}\Gamma ((D+1)/2)}{\sqrt{\pi }\left( b^{2}+\lambda ^{2}\right) ^{(D+1)/2}}%
,  \label{IntBes}
\end{equation}%
to the leading order we get%
\begin{equation}
\int_{0}^{\infty }dy\,f(y)e^{-by}w_{D/2}(\lambda y)\approx \frac{2^{D/2}}{%
\sqrt{\pi }\lambda }\Gamma ((D+1)/2)f(0).  \label{IntAs}
\end{equation}%
For the integral in (\ref{jlsmalla2}) we take $b=2$ and%
\begin{equation}
f(y)=\frac{2+\sum_{j=1,2}e^{-2yz_{j}/a}/c_{j}(y)}{c_{1}(y)c_{2}(y)-e^{-2y}}.
\label{fy}
\end{equation}%
In the case of non-Neumann boundary conditions one has $f(0)=1$ and, hence,%
\begin{equation}
\Delta \langle j^{l}\rangle \approx \frac{2e\Gamma ((D+1)/2)}{\pi
^{(D+1)/2}L^{D}}\sum_{n=1}^{\infty }\frac{\sin \left( n\tilde{\alpha}%
_{l}\right) }{n^{D}}.  \label{jlsmalla3}
\end{equation}%
Combining this result with the expressions from the previous section for the
geometry of a single plate, we conclude that $\lim_{a\rightarrow 0}\langle
j^{l}\rangle =0$, i.e., for non-Neumann boundary conditions the total
current density in the region between the plates tends to zero for small
separations between the plates. For non-Neumann boundary condition on one
plate and Neumann boundary condition on the other we have $f(0)=-1$ and the
corresponding formula is obtained from (\ref{jlsmalla3}) changing the sign
of the right-hand side. In this case we have again $\lim_{a\rightarrow
0}\langle j^{l}\rangle =0$.

\begin{figure}[tbph]
\begin{center}
\epsfig{figure=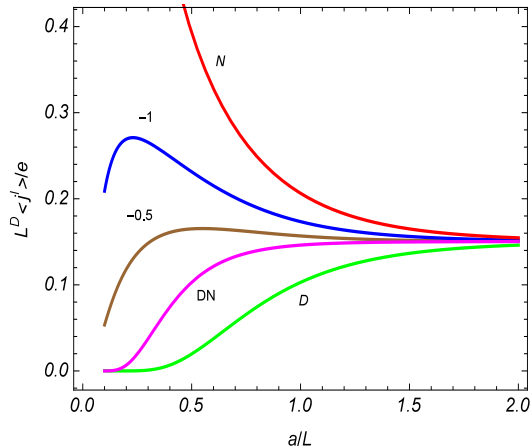,width=7.cm,height=6.cm}
\end{center}
\caption{The VEV of the current density in the region between the plates
evaluated at $z=a/2$, as a function of the separation between the plates.
The graphs are plotted for Dirchlet and Neumann boundary conditions on both
plates, for Dirichlet condition on the left plate and Neumann condition on
the right one, and for Robin boundary conditions with the values of $\protect%
\beta _{j}/L$ given near the curves. For the phase we have taken the value $%
\tilde{\protect\alpha}=\protect\pi /2$.}
\label{fig3}
\end{figure}

For Neumann boundary condition on both plates, for the function in (\ref{fy}%
) we have $f(y)\sim 2/y$, $y\rightarrow 0$. In order to obtain the leading
term in the asymptotic expansion for small values of $a$ it is more
convenient to use the expression (\ref{jlDNaltm0}) with the lower sign
instead of the right-hand side of (\ref{jlsmalla2}). For small $a/L$ the
dominant contribution in (\ref{jlDNaltm0}) comes from large values of $r$
and, to the leading order, we replace the corresponding summation by the
integration. For the leading term this gives
\begin{equation}
\langle j^{l}\rangle \approx \frac{2e\Gamma (D/2)}{\pi ^{D/2}L^{D-1}a}%
\sum_{n=1}^{\infty }\frac{\sin \left( n\tilde{\alpha}\right) }{n^{D-1}},
\label{jlsmalla4}
\end{equation}%
and for Neumann boundary condition the current density diverges in the limit
$a\rightarrow 0$ like $1/a$. The described features in the behavior of the
vacuum current density, $L^{D}\langle j^{l}\rangle /e$, in the region
between the plates located at $z=0$ and $z=a$, as a function of the
separation between the plates, is illustrated in figure \ref{fig3} for a $D=4
$ massless scalar field in the model with a single compact dimension of the
length $L$ and of the phase $\tilde{\alpha}$. The graphs are plotted for $%
z=a/2$ and $\tilde{\alpha}=\pi /2$, in the cases of Dierichlet (D), Neumann
(N) boundary conditions on both plates, for Dirichlet boundary condition at $%
z=0$ and Neumann boundary condition at $z=a$ (DN), and for Robin boundary
conditions with $\beta _{j}/L=-0.5$ and $\beta _{j}/L=-1$ (numbers near the
curves). At large separations between the plates, the boundary-induced
effects are small and the current density coincides with that in the
boundary-free geometry.

\begin{figure}[tbph]
\begin{center}
\begin{tabular}{cc}
\epsfig{figure=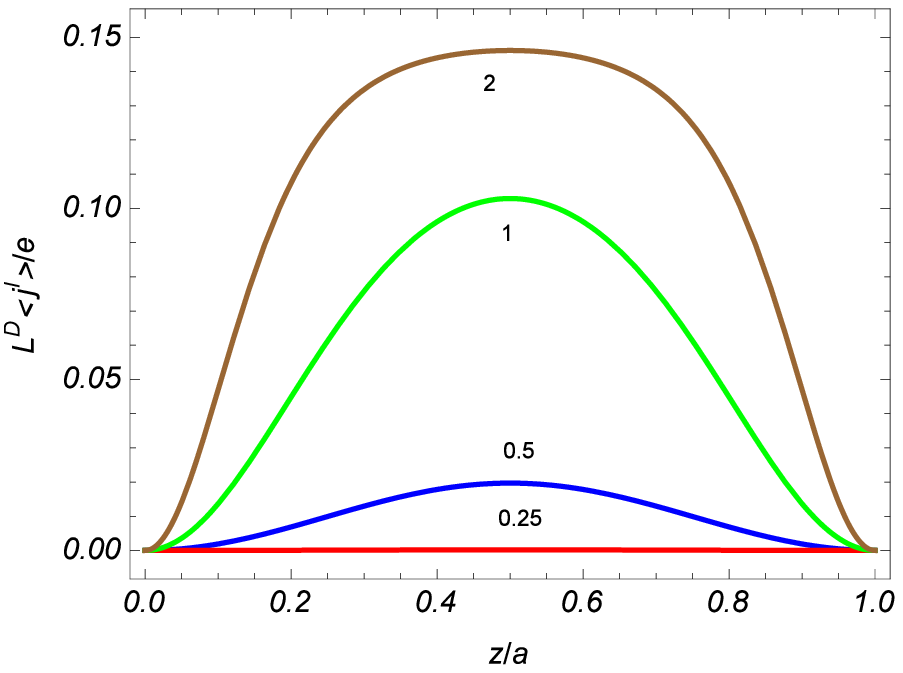,width=7.cm,height=6.cm} & \quad %
\epsfig{figure=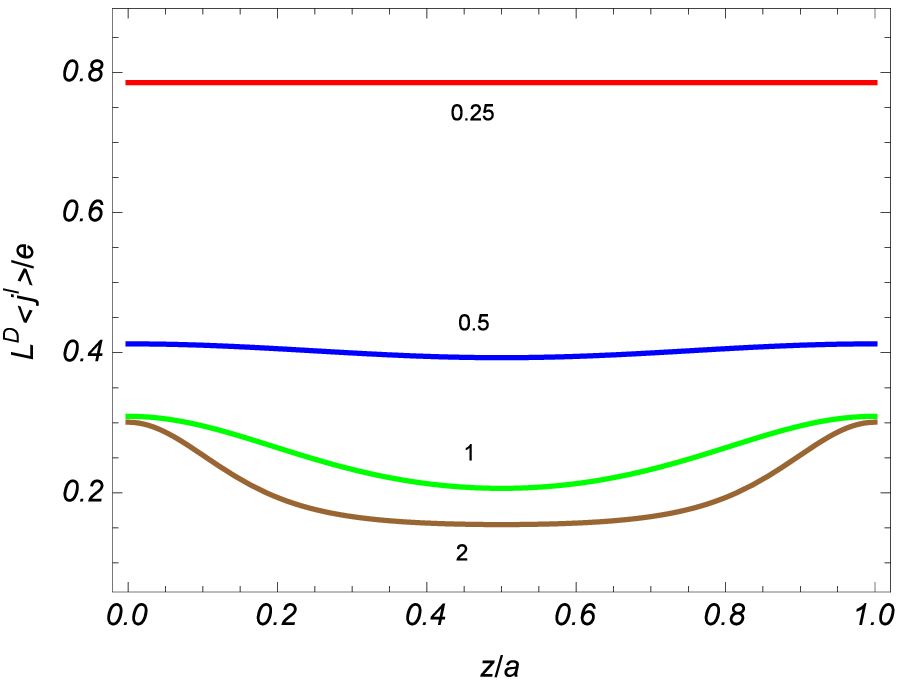,width=7.cm,height=6cm}%
\end{tabular}%
\end{center}
\caption{The current density between the plates as a function of the
relative distance from the left plate in the model with a single compact
dimension. The graphs are plotted for a massless field with the parameter $%
\tilde{\protect\alpha}=\protect\pi /2$ and with Dirichlet (left panel) and
Neumann (right panel) boundary conditions. The numbers near the curves
correspond to the values of $a/L$. }
\label{fig4}
\end{figure}

In figure \ref{fig4}, in the model with a single compact dimension of the
length $L$ and for a $D=4$ massless scalar field with Dirichlet (left panel)
and Neumann (right panel) boundary conditions, we have plotted the total
current density as a function of the ratio $z/a$ in the region between the
plates. The numbers near the curves correspond to the values of $a/L$ and
the graphs are plotted for $\tilde{\alpha}=\pi /2$. The features, obtained
before on the base of asymptotic analysis, are clearly seen from the graphs:
the current density for Dirichlet/Neumann scalar decreases/increases with
decreasing separation between the plates and for Dirichlet scalar it
vanishes on the plates.

\begin{figure}[tbph]
\begin{center}
\begin{tabular}{cc}
\epsfig{figure=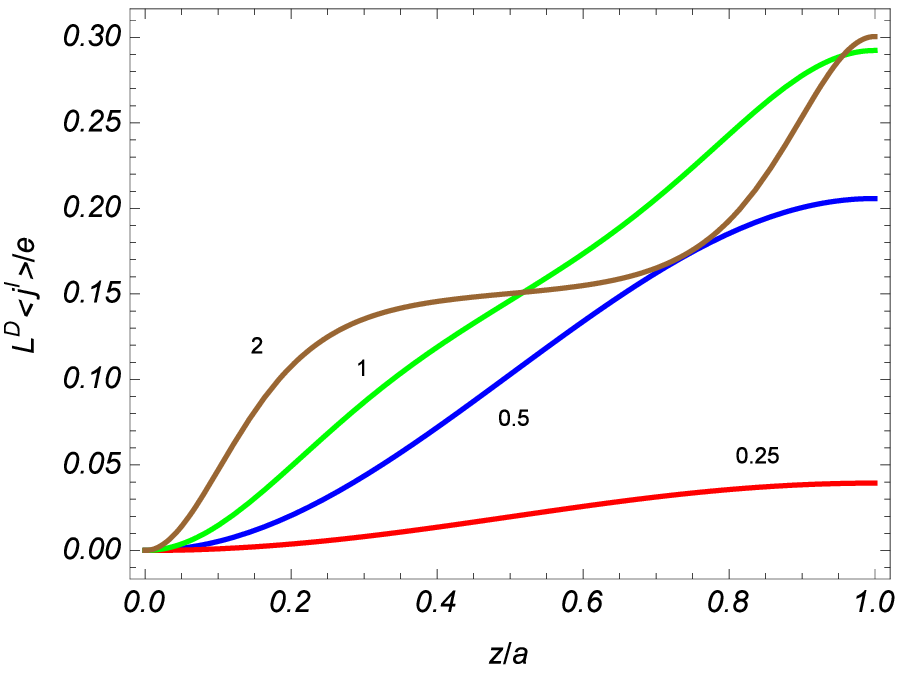,width=7.cm,height=6.cm} & \quad %
\epsfig{figure=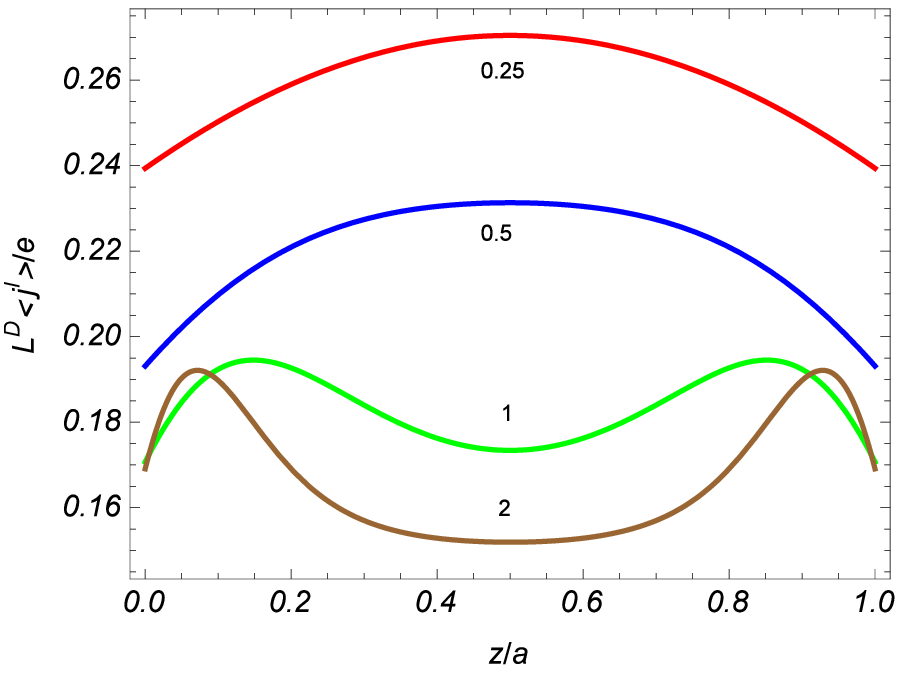,width=7.cm,height=6cm}%
\end{tabular}%
\end{center}
\caption{The same as in figure \protect\ref{fig4} for Dirichlet boundary
condition on the left plate and Neumann condition on the right one (left
panel). The right panel is plotted for Robin boundary condition on both
plates with $\protect\beta _{1}/L=\protect\beta _{2}/L=-1$. }
\label{fig5}
\end{figure}

The same graphs for Dirichlet boundary condition on the left plate and
Neumann condition on the right one are presented on the left panel of figure %
\ref{fig5}. The right panel in figure \ref{fig5} is plotted for Robin
boundary condition on both plates with $\beta _{1}/L=\beta _{2}/L=-1$. In
the Robin case, the current density decreases with the further decrease of
the separation between the plates and it tends to zero in the limit $%
a\rightarrow 0$, in accordance with the general analysis described above.

\section{Conclusion}

\label{Sec:Conc}

In the present paper we have investigated the influence of parallel flat
boundaries on the VEV of the current density for a charged scalar field in a
flat spacetime with toroidally compactified spatial dimensions, assuming the
presence of a constant gauge field. The effect of the latter on the current
is similar to the Aharonov-Bohm effect and is caused by the nontrivial
topology of the background space. Along compact dimensions we have
considered quasiperiodicity conditions with general phases. The special
cases of twisted and untwisted fields are the configurations most frequently
discussed in the literature. By a gauge transformation, the problem with a
constant gauge field is mapped to the one with zero field, shifting the
phases in the periodicity conditions by an amount proportional to the
magnetic flux enclosed by a compact dimension in the initial representation
of the model. On the plates we employed Robin boundary conditions, in
general, with different coefficients on the left and right plates. The Robin
boundary conditions for bulk fields naturally arise in braneworld scenario
and the boundaries considered here may serve as a simple model for the
branes.

We considered a free field theory and all the information on the properties
of the vacuum state is encoded in two-point functions. Here we chose the
Hadamard function. The VEV of the current density is obtained from this
function in the coincidence limit by using (\ref{jl1}). For the evaluation
of the Hadamard function we have employed a direct summation over the
complete set of modes. In the region between the plates the eigenvalues of
the momentum component perpendicular to the plates are quantized by the
boundary conditions on the plates and are given implicitly, in terms of
solutions of the transcendental equation (\ref{EigEq2}). Depending on the
values of the Robin coefficients, this equation may have purely imaginary
solutions $y=\pm iy_{l}$. In order to have a stable vacuum with $\langle
\varphi \rangle =0$, we assume that $\omega _{0}>y_{l}$. Compared to the
case of the bulk with trivial topology, this constraint in models with
compact dimensions is less restrictive. The eigenvalues of the momentum
components along compact dimensions are quantized by the periodicity
conditions and are determined by (\ref{kltild}). The application of the
generalized Abel-Plana formula for the summation over the roots of (\ref%
{EigEq2}) allowed us to extract from the Hadamard function the part
corresponding to the geometry with a single plate and to present the
second-plate-induced contribution in the form which does not require the
explicit knowledge of the eigenmodes for $k_{p+1}$ (see (\ref{G(1)})). In
addition, the corresponding integrand decays exponentially in the upper
limit. A similar representation, (\ref{G1jb}), is obtained for the Hadamard
function in the geometry of a single plate. The second term in the
right-hand side of this representation is the boundary-induced contribution.
An alternative representation for the Hadamard function, (\ref{G1alt3}), is
obtained in Appendix, by making use of the summation formula (\ref{AbelPlan1}%
). The second term in the right-hand side of this representation is the
contribution induced by the compactification of the $l$th dimension.

The VEVs of the charge density and the components of the current density
along uncompact dimensions vanish. The current density along compact
dimensions is a periodic function of the magnetic flux with the period equal
to the flux quantum. The component along the $l$th compact dimension is an
odd function of the phase $\tilde{\alpha}_{l}$ and an even function of the
remaining phases $\tilde{\alpha}_{i}$, $i\neq l$. First we have considered
the geometry with a single plate. The VEV of the current density is
decomposed into the boundary-free and plate-induced parts. The boundary-free
contribution was investigated in \cite{Beze13} and we have been mainly
concerned with the plate-induced part, given by (\ref{jlj(1)}). For special
cases of Dirichlet and Neumann boundary conditions the corresponding
expression is simplified to (\ref{jlj(1)DN}). The plate-induced part has
opposite signs for Dirichlet and Neumann conditions. At distances from the
plate larger than the lengths of compact dimensions the asymptotic is
described by (\ref{jlj(1)far}) and the plate-induced contribution is
exponentially small. For the investigation of the near-plate asymptotic of
the current density it is more convenient to use the representation (\ref%
{jl1alt}) for the general Robin case and (\ref{jl1altDN}) for Dirichlet and
Neumann conditions. From these representations it follows that the current
density is finite on the plate. This property is in sharp contrast with the
behavior of the VEVs of the field squared and of the energy-momentum tensor
which diverge on the plate. For Dirichlet boundary condition the current
density vanishes on the plate and for Neumann condition its value on the
plate is two times larger than the current density in the boundary-free
geometry. The normal derivative of the current density vanishes on the plate
for both Dirichlet and Neumann conditions. This is not the case for general
Robin condition. The behavior of the plate-induced part of the current
density along $l$th dimension, in the limit when the lengths of the other
compact dimensions are much smaller than $L_{l}$, crucially depend wether
the phases $\tilde{\alpha}_{i}$, $i\neq l$, are zero or not. For $%
\sum\nolimits_{i\neq l}\tilde{\alpha}_{i}^{2}\neq 0$ one has $\omega
_{0l}\neq 0$ and the corresponding asymptotic expression is given by (\ref%
{jl1SmalLr}). In this case the plate-induced contribution is exponentially
suppressed. For $\tilde{\alpha}_{i}=0$, $i\neq l$, the leading term in the
asymptotic expansion, multiplied by $V_{q}/L_{l}$, coincides with the
corresponding current density for $(p+2)$-dimensional space with topology $%
R^{p+1}\times S^{1}$. In the limit when the length of the $l$th dimension is
much larger than the other length scales of the model, the behavior of the
plate-induced contribution to the current density is essentially different
for the cases $\omega _{0l}\neq 0$ and $\omega _{0l}=0$. In the former case
the leading term is given by (\ref{jlLlarge}) and the current density is
suppressed by the factor $e^{-L_{l}\omega _{0l}}$. In the second case, for
the leading term one has the expression (\ref{jlLlargeb}) and its behavior,
as a function of $L_{l}$, is power law. In both cases and for non-Neumann
boundary conditions, the leading terms in the boundary-induced and
boundary-free parts of the current density cancel each other.

For the current density in the region between the plates we have provided
various decompositions ((\ref{jl2}), (\ref{jldec}), (\ref{jltot}) for
general Robin boundary conditions and (\ref{jl2DN}), (\ref{jl2DN2}), (\ref%
{jlDN}) for special cases of Dirichlet and Neumann conditions). In the case
of Dirichlet boundary condition the total current vanishes on the plates.
The normal derivative vanishes on the plates for both Dirichlet and Neumann
cases. In the limit when the separation between the plates is smaller than
all the length scales in the problem, the behavior of the current density is
essentially different for non-Neumann and Neumann boundary conditions. In
the former case, the total current density in the region between the plates
tends to zero. For Neumann boundary condition on both plates, for small
separations the total current density is dominated by the interference part
and it diverges inversely proportional to the separation (see (\ref%
{jlsmalla4})). The results of the present paper may be applied to
Kaluza-Klein-type models in the presence of branes (for $D>3$) and to planar
condensed matter systems (for $D=2$), described within the framework of an
effective field theory. In particular, in the former case, the vacuum
currents along compact dimensions generate magnetic fields in the
uncompactified subspace. The boundaries discussed above can serve as a
simple model for the edges of planar systems.

\section{Acknowledgments}

N. A. S. was supported by the State Committee of Science Ministry of
Education and Science RA, within the frame of Research Project No. 15 RF-009.

\appendix

\section{Alternative representation of the Hadamard function}

\label{sec:App}

In this section we derive an alternative representation for the Hadamard
function which is well suited for the investigation of the near-plate
asymptotic of the current density. The starting point is the representation (%
\ref{G11}). We apply to the corresponding series over $n_{l}$ the summation
formula \cite{Bell10,Beze08}%
\begin{eqnarray}
&&\frac{2\pi }{L_{l}}\sum_{n_{l}=-\infty }^{\infty
}g(k_{l})f(|k_{l}|)=\int_{0}^{\infty }du[g(u)+g(-u)]f(u)  \notag \\
&&\qquad +i\int_{0}^{\infty }du\,[f(iu)-f(-iu)]\sum_{\lambda =\pm 1}\frac{%
g(i\lambda u)}{e^{uL_{l}+i\lambda \tilde{\alpha}_{l}}-1},  \label{AbelPlan1}
\end{eqnarray}%
where $k_{l}$ is given by (\ref{kltild}). The part in the Hadamard function
coming from the first term in the right-hand side of (\ref{AbelPlan1})
coincides with the Hadamard function for the geometry of two plates in $D$%
-dimensional space with topology $R^{p+2}\times T^{q-1}$ and with the
lengths of the compact dimensions $(L_{p+2},\ldots ,L_{l-1},L_{l+1},\ldots
,L_{D})$ (the $l$th dimension is uncompactified). We will denote this
function by $G_{R^{p+2}\times T^{q-1}}(x,x^{\prime })$. As a result, under
the assumption $\beta _{j}\leqslant 0$, the Hadamard function is decomposed
as%
\begin{eqnarray}
G(x,x^{\prime }) &=&G_{R^{p+2}\times T^{q-1}}(x,x^{\prime })+\frac{L_{l}}{%
\pi aV_{q}}\int \frac{d\mathbf{k}_{p}}{(2\pi )^{p}}\sum_{\mathbf{n}_{q-1}}
\notag \\
&&\times \sum_{n=1}^{\infty }\frac{\lambda _{n}g(z,z^{\prime },\lambda
_{n}/a)e^{i\mathbf{k}_{p}\cdot \Delta \mathbf{x}_{p}+i\mathbf{k}%
_{q-1}^{l}\cdot \Delta \mathbf{x}_{q-1}^{l}}}{\lambda _{n}+\cos \left[
\lambda _{n}+2\tilde{\gamma}_{j}(\lambda _{n})\right] \sin \lambda _{n}}
\notag \\
&&\times \int_{\omega _{\mathbf{k}}^{(l)}}^{\infty }du\,\frac{\cosh (\Delta t%
\sqrt{u^{2}-\omega _{\mathbf{k}}^{(l)2}})}{\sqrt{u^{2}-\omega _{\mathbf{k}%
}^{(l)2}}}\sum_{\lambda =\pm 1}\frac{e^{-\lambda u\Delta x^{l}}}{%
e^{uL_{l}+i\lambda \tilde{\alpha}_{l}}-1},  \label{G1alt}
\end{eqnarray}%
where $\mathbf{x}_{q-1}^{l}=(x^{p+2},...,x^{l-1},x^{l+1},\ldots x^{D})$, $%
\mathbf{k}_{q-1}=(k_{p+2},\ldots ,k_{l-1},k_{l+1},\ldots ,k_{D})$, and $%
\omega _{\mathbf{k}}^{(l)}=\sqrt{\omega _{\mathbf{k}}^{2}-k_{l}^{2}}$. Here,
the second term in the right-hand side vanishes in the limit $%
L_{l}\rightarrow \infty $ and is induced by the compactification of the $l$%
th dimension from $R^{1}$ to $S^{1}$ with the length $L_{l}$.

By making use of the relation%
\begin{equation}
\sum_{\lambda =\pm 1}\frac{e^{-\lambda u\Delta x^{l}}}{e^{uL_{l}+i\lambda
\tilde{\alpha}_{l}}-1}=2u\sum_{r=1}^{\infty }h_{r}(u,\Delta x^{l}),
\label{Rel2}
\end{equation}%
with%
\begin{equation}
h_{r}(\Delta x^{l},u)=\frac{e^{-ruL_{l}}}{u}\cosh \left( u\Delta x^{l}+ir%
\tilde{\alpha}_{l}\right) ,  \label{hl}
\end{equation}%
we rewrite the formula (\ref{G1alt}) in the form%
\begin{eqnarray}
G(x,x^{\prime }) &=&G_{R^{p+2}\times T^{q-1}}(x,x^{\prime })+\frac{2L_{l}}{%
\pi aV_{q}}\sum_{r=1}^{\infty }\int \frac{d\mathbf{k}_{p}}{(2\pi )^{p}}
\notag \\
&&\times \sum_{\mathbf{n}_{q-1}}\int_{0}^{\infty }dy\,\cosh (y\Delta t)e^{i%
\mathbf{k}_{p}\cdot \Delta \mathbf{x}_{p}+i\mathbf{k}_{q-1}^{l}\cdot \Delta
\mathbf{x}_{q-1}^{l}}  \notag \\
&&\times \sum_{n=1}^{\infty }\frac{\lambda _{n}g(z,z^{\prime },\lambda
_{n}/a)h_{r}(\Delta x^{l},\sqrt{\lambda _{n}^{2}/a^{2}+y^{2}+\omega _{p,%
\mathbf{n}_{q-1}}^{2}})}{\lambda _{n}+\cos \left[ \lambda _{n}+2\tilde{\gamma%
}_{j}(\lambda _{n})\right] \sin \lambda _{n}},  \label{G1alt2}
\end{eqnarray}%
with $\omega _{p,\mathbf{n}_{q-1}}=\sqrt{\mathbf{k}_{p}^{2}+\omega _{\mathbf{%
n}_{q-1}}^{2}}$. Now, by using the summation formula (\ref{sumfor}) for the
series over $n$ we get the final representation%
\begin{eqnarray}
G(x,x^{\prime }) &=&G_{R^{p+2}\times T^{q-1}}(x,x^{\prime })+\frac{2L_{l}}{%
\pi ^{2}V_{q}}\sum_{r=1}^{\infty }\int \frac{d\mathbf{k}_{p}}{(2\pi )^{p}}%
\sum_{\mathbf{n}_{q-1}}e^{i\mathbf{k}_{p}\cdot \Delta \mathbf{x}_{p}+i%
\mathbf{k}_{q-1}^{l}\cdot \Delta \mathbf{x}_{q-1}^{l}}  \notag \\
&&\times \int_{0}^{\infty }dy\,\cosh (\Delta ty)\Big\{\int_{0}^{\infty
}dug_{j}(z,z^{\prime },u)h_{r}(\Delta x^{l},\sqrt{u^{2}+y^{2}+\omega _{p,%
\mathbf{n}_{q-1}}^{2}})  \notag \\
&&+\int_{\sqrt{y^{2}+\omega _{p,\mathbf{n}_{q-1}}^{2}}}^{\infty }du\frac{%
g_{j}(z,z^{\prime },iu)}{c_{1}(au)c_{2}(au)e^{2au}-1}\sum_{s=\pm
1}ih_{sr}(\Delta x^{l},i\sqrt{u^{2}-y^{2}-\omega _{p,\mathbf{n}_{q-1}}^{2}})%
\Big\}.  \label{G1alt3}
\end{eqnarray}%
In this expression, the part with the first term in the figure braces is the
contribution to the Hadamard function induced by the compactification of the
$l$th dimension for the geometry of a single plate at $x^{p+1}=a_{j}$ and
the part with the second term in the figure braces is induced by the second
plate. Note that the contribution of the first term in the right-hand side
of (\ref{G1alt3}) to current density along the $l$th dimension vanishes.

In deriving the representation (\ref{G1alt3}) we have assumed that $\beta
_{j}\leqslant 0$. For this case, in the region between the plates, all the
eigenvalues for the momentum $k_{p+1}$ are real and in the geometry of a
single plate there are no bound states. For $\beta _{j}>0$, in the
application of the summation formula (\ref{sumfor}) to the series over $n$
in (\ref{G1alt2}) the contribution from the poles $\pm i/b_{j}$ should be
added to the right-hand side of (\ref{sumfor}). This contribution comes from
the bound state in the geometry of a single plate at $x^{p+1}=a_{j}$. For
this bound state the mode function has the form $\varphi _{\mathbf{k}}^{(\pm
)}(x)\sim e^{-z_{j}/\beta _{j}}e^{i\mathbf{k}_{\parallel }\cdot \mathbf{x}%
_{\parallel }\mp i\omega _{\mathbf{k}}^{(b)}t}$ with $\omega _{\mathbf{k}%
}^{(b)}=\sqrt{\mathbf{k}_{p}^{2}+\omega _{\mathbf{n}_{q}}^{2}-1/\beta
_{j}^{2}}$. Assuming that $\omega _{0l}>1/\beta _{j}$, the contribution from
the bound state to the Hadamard function in the geometry of a single plate
is given by the expression%
\begin{eqnarray}
G_{bj}^{(1)}(x,x^{\prime }) &=&\frac{4\theta (\beta _{j})L_{l}}{\pi
V_{q}\beta _{j}}e^{-|z+z^{\prime }-2a_{j}|/\beta _{j}}\sum_{r=1}^{\infty
}\int \frac{d\mathbf{k}_{p}}{(2\pi )^{p}}\sum_{\mathbf{n}_{q-1}^{l}}%
\int_{0}^{\infty }dx\,e^{i\mathbf{k}_{p}\cdot \Delta \mathbf{x}_{p}+i\mathbf{%
k}_{q-1}^{l}\cdot \Delta \mathbf{x}_{q-1}^{l}}  \notag \\
&&\times \cosh (x\Delta t)h_{r}(\Delta x^{l},\sqrt{x^{2}+\mathbf{k}%
_{p}^{2}+\omega _{\mathbf{n}_{q-1}}^{2}-1/\beta _{j}^{2}}),  \label{Gjbound}
\end{eqnarray}
where $\theta (x)$ is the Heaviside unit step function. In the case $\omega
_{0l}<1/\beta _{j}<\omega _{0}$ the corresponding expression is more
complicated.

\end{document}